\documentclass{aa}

\usepackage[T1]{fontenc}
\usepackage[utf8]{inputenc}
\usepackage[final]{microtype}
\usepackage[varg]{txfonts}

\usepackage[english]{babel}
\usepackage[intlimits]{mathtools}
\usepackage{subcaption}
\usepackage{amssymb}
\usepackage{gensymb}
\usepackage{bm }
\usepackage[separate-uncertainty,table-align-uncertainty]{siunitx}

\usepackage{natbib}
\bibpunct{(}{)}{;}{a}{}{,} 
\usepackage[inline]{enumitem}
\usepackage[normalem]{ulem}

\usepackage{graphicx} 
\graphicspath{{../figures/}}

\usepackage{hyperref}
\hypersetup{colorlinks=true,       
        linkcolor=red,          
        citecolor=blue,        
        filecolor=magenta,      
        urlcolor=cyan,           
}
\usepackage{cleveref}

\newcommand{\YCC}{\textit{C15}}  

\newcounter{algo}[section]
\newtheorem{alg}[proof]{Algorithm}{\bf}{\refstepcounter{algo}\rm}

\crefname{alg}{algorithm}{algorithms}

\begin{document}

\title{A novel cosmic filament catalogue from SDSS data%
	\thanks{The catalogue is available in electronic form at \href{https://www.javiercarron.com/catalogue}{https://www.javiercarron.com/catalogue} or at the CDS via anonymous ftp to cdsarc.u-strasbg.fr (130.79.128.5)	or via \href{http://cdsweb.u-strasbg.fr/cgi-bin/qcat?J/A+A/}{http://cdsweb.u-strasbg.fr/cgi-bin/qcat?J/A+A/} }}
\author{Javier Carrón Duque\inst{1,2}\thanks{Email: \href{mailto:javier.carron@roma2.infn.it}{javier.carron@roma2.infn.it}} %
        \and Marina Migliaccio\inst{1,2} %
        \and Domenico Marinucci\inst{3} %
        \and Nicola Vittorio\inst{1,2}
}
\institute{Dipartimento di Fisica, Universit\`{a} di Roma Tor Vergata, via della Ricerca Scientifica, 1, 00133, Roma, Italy %
        \and INFN, Sezione di Roma 2, Universit\`{a} di Roma Tor Vergata, via della Ricerca Scientifica, 1, 00133 Roma, Italy %
        \and Dipartimento di Matematica, Università di Roma Tor Vergata, via della Ricerca Scientifica, 1, 00133, Roma, Italy}
\date{}

\abstract{}%
{In this work we present a new catalogue of cosmic filaments obtained from the latest Sloan Digital Sky Survey (SDSS) public data.}%
{In order to detect filaments, we implement a version of the Subspace-Constrained Mean-Shift algorithm that is boosted by machine learning techniques. This allows us to detect cosmic filaments as one-dimensional maxima in the galaxy density distribution. Our filament catalogue uses the cosmological sample of SDSS, including Data Release 16, and therefore inherits its sky footprint (aside from small border effects) and redshift coverage. In particular, this means that, taking advantage of the quasar sample, our filament reconstruction covers redshifts up to $z=2.2$, making it one of the deepest filament reconstructions to our knowledge. We follow a tomographic approach and slice the galaxy data in $269$ shells at different redshift. The reconstruction algorithm is applied to $2D$ spherical maps.}%
{The catalogue provides the position and uncertainty of each detection for each redshift slice. The quality of our detections, which we assess with several metrics, show improvement with respect to previous public catalogues obtained with similar methods. We also detect a highly significant correlation between our filament catalogue and galaxy cluster catalogues built from microwave observations of the Planck Satellite and the Atacama Cosmology Telescope.}%
{}


        \maketitle
        
        \section{Introduction}
        
        It has long been recognised that matter in the Universe is arranged along an intricate pattern known as the cosmic web \citep{joeveer1978,bond1996}. This highly structured web results from the anisotropic gravitational collapse and hosts four main classes of substructures: massive halos, long filaments, sheets, and vast low-density regions known as cosmic voids \citep{zeldovich1982}. Filaments are highly defining features in this network, as they delimit voids and provide bridges along which matter is expected to be channelled towards halos and galaxy clusters. 
        
        Filamentary structures in the large-scale distribution of galaxies were originally observed with the Center for Astrophysics (CfA) galaxy survey \citep{delapparent1986}. Since then, ample evidence for such structures has been gathered by galaxy redshift surveys, such as the Sloan Digital Sky Survey (SDSS, \citealp{york2000}), the Two-degree-Field Galaxy Redshift Survey (2dFGRS, \citealp{cole2005}), the Six-degree-Field Galaxy Survey (6dFGS, \citealp{jones2004}), the Cosmic Evolution Survey (COSMOS, \citealp{scoville2007}), the Galaxy and Mass Assembly (GAMA, \citealp{driver2011}), the Two Micron All Sky Survey (2MASS, \citealp{huchra2012}), and the VIMOS Public Extragalactic Redshift Survey (VIPERS, \citealp{guzzo2014}). Concurrently, further corroboration of the filamentary arrangement of matter also came from cosmological N-body simulations ({\it e.g.}, \citealp{springel2005,vogelsberger2014}). 
        
        Nevertheless, detecting and reconstructing cosmic filaments  still poses some challenges due to the lack of a standard and unique definition. As a consequence, a variety of approaches and algorithms have been explored in the literature. Among these there are multiscale morphology filters, such as MMF (\citealp{aragoncalvo2007,aragoncalvo2010}) or NEXUS (\citealp{cautun2013}); segmentation-based approaches, such as the Candy model (\citealp{stoica2005}) or the watershed method implemented in SpineWeb (\citealt{aragoncalvo2010a}); skeleton analyses (\citealp{novikov2006,sousbie2008}); tessellations, as in DisPerSE (\citealp{sousbie2011}); the Bisous model (\citealp{tempel2016}); methods based on graph theory such as T-Rex (\citealp{bonnaire2019}) or minimal spanning trees (\citealp{pereyra2019}); or methods based on machine learning classification algorithms, such as random forests (\citealp{buncher2020}) or convolutional neural networks (CNNs; as in \citealp{aragoncalvo2019}). Comparisons between different methods can be found in \citet{libeskind2017} and \citet{rost2020}, for example. All these methods rely on different assumptions and can be more or less advantageous depending on the specific application. 
        
        In the present paper, we adopt and extend the technique proposed by \citet{chen2015} (hereafter \YCC{}), who implement the Subspace-Constrained Mean-Shift (SCMS) algorithm to identify the filamentary structure in the distribution of galaxies. This technique is based on a modified gradient ascent method that models filaments as ridges in the galaxy density distribution. It has already been successfully applied to SDSS data \citep{chen2016}, providing a filament catalogue which has been used to study the effect of filaments on galaxy properties, such as colour and mass \citep{chen2017}, or orientation \citep{chen2019,krolewski2019}. The catalogue also allowed the first detection of cosmic microwave background (CMB) lensing by cosmic filaments \citep{he2018} and has been used to study the possible effect of cosmic strings on cosmic filaments \citep{fernandez2020}. Our implementation of the algorithm presents two main differences with respect to \YCC{}. In their work, the filament finder is applied on 2D galaxy density maps which are built from the original catalogue assuming a flat-sky approximation. Here, we extend the formalism to work on spherical coordinates, as we expect this approach to be less sensitive to projection effects, especially for wide surveys. Secondly, we complement the method with a machine learning technique that combines information from different smoothing scales in the density field. This procedure has been designed to be robust against the choice of a particular smoothing scale, and allows the algorithm to exploit the information present at lower redshift (with more available galaxies) in order to robustly predict the filaments at higher redshift (with fewer available galaxies).
        
        We apply this new implementation of the algorithm to the DR16 SDSS data \citep{ahumada2020} from BOSS (Baryon Oscillation Spectroscopic Survey) and eBOSS (extended BOSS) clustering surveys. The latter, in particular, has not been used in previous filament-extraction studies. eBOSS creates the largest volume survey to date and, thanks to its sample of quasars, we have been able to build a filament catalogue that extends to very high redshifts of up to $z=2.2$. In our approach, we divide the data into $269$ redshift shells of $20$ Mpc in  width, and provide the filament reconstruction in 2D spherical slices. We also deliver the uncertainties associated to the filament positions, which have been estimated with a bootstrap method. 
        
        There are several advantages to using this tomographic approach to filament reconstruction. For example, the 2D filament maps can be directly employed in cross-correlation studies with maps of the CMB lensing convergence or thermal Sunyaev-Zel'dovich signal (see {\it e.g.}, \citealt{tanimura2020}). We also expect this method to be particularly suitable for the extraction of filament maps from future wide photometric surveys, such as Euclid \citep{laureijs2011} and the Vera C. Rubin Observatory Legacy Survey of Space and Time (LSST, \citealp{ivezic2019}), where uncertainties on redshift estimates might limit the applicability of $3D$ methods. The tomographic approach has also been shown to provide more robust detection near galaxy clusters and to be less sensitive to the Finger of God effect, even when it is statistically corrected \citep{kuchner2021}. Moreover, the tomographic catalogue will naturally trace the evolution of the filamentary structure of the Universe  as a function of redshift and can therefore\textbf{ }be used to test structure formation history and cosmological models. On the other hand, the tomographic approach reduces the sensitivity to filaments at angles close to the line of sight and may split long filaments across two or more slices. Different applications will benefit from different approaches.
        
        The paper is organised as follows. We start by providing a detailed description of the methodology and the algorithm in \Cref{s:methodology}. In \Cref{s:data} we introduce the data sets used in our analysis, while in \Cref{s:results} we present the results of the filament extraction. In \Cref{s:comparison} we compare the results with the catalogue from \YCC{}. In \Cref{s:validation}
        we discuss the validation of the catalogue through several metrics. Finally, in \Cref{s:Conclusions} we draw our conclusions.        
        
        \section{Methodology}
        \label{s:methodology}
        
        In this section, we explain the theoretical framework that we use and the complete methodology that we implement. The scheme of this section is as follows. In \Cref{ss:definition} we explain our working definition of cosmic filaments. In \Cref{ss:tomo} we introduce our tomographic strategy. In \Cref{ss:scms} we explain the basis of the SCMS algorithm used to detect these filaments from galaxy distribution maps. In \Cref{ss:err} we study how this algorithm can be extended to yield an estimation of the error of the detection. In \Cref{ss:size} we introduce the two methods (based on SCMS) that we implement to detect filaments, including a version of SCMS boosted by machine learning techniques. Finally, in \Cref{ss:pros} we discuss the strengths and limitations of our methodology. All the details explaining the training of the machine learning step and the procedure to clean the final filament catalogue and remove outliers can be found in \Cref{ap:train} and \Cref{ap:clean}, respectively.
        
        \subsection{Definition of cosmic filaments}
        \label{ss:definition}
        
        Cosmic filaments can be defined in multiple ways. In this work, we adopt the mathematical framework of the \textit{Ridge formalism} \citep{eberly1996}. This formalism has been widely studied in the mathematical literature, for example by \citet{ozertem2011} and \citet{genovese2014}. We refer the reader to these references for a more complete overview of the properties of ridges. Here, we summarise the definition and some key properties.
        
        A ridge of a density field $d(x)$ in $D$ dimensions is defined as follows. Let $g(x)$ and $H(x)$ be the gradient and the hessian of $d(x)$. Let $\lambda_1\geq...\geq\lambda_D$ be the eigenvalues of $H(x)$, with associated eigenvectors $v_1,...,v_D$. We define $V(x)$ as the subspace spanned by all eigenvectors except $v_1$ (therefore perpendicular to it). We define the gradient projected into this subspace as $G(x)$. Finally, the ridges of the density field $d(x)$ are defined as 
        \begin{equation}
                \text{Ridge}(d) \equiv \{x\ |\ G(x)=0,\ \lambda_2<0\}.
        \end{equation}
        
        This definition is conceptually analogous to the criteria for maxima of a one-dimensional function: `first derivative equals $0$' and `second derivative is negative'. In the case of higher dimensions, one has to restrict those criteria to the subspace perpendicular to the filament. The largest eigenvalue $\lambda_1$ corresponds to an eigenvector $v_1$ parallel to the filament, given that the density decreases with the distance to a filament, but not in the direction of the filament (or not as quickly). We note that the gradient on a point of the ridge is always parallel to the direction of the ridge. Not only is this observation  important for the algorithm we use to find the filaments, but it also provides a method for calculating the direction of a filament in a point.
        
        By definition, ridges are one-dimensional, overdense structures, the two properties that define cosmic filaments. We note that other cosmological structures are given by slightly modified definitions. If the subspace $V(x)$ is the whole space and $\lambda_1<0$, we obtain overdense zero-dimensional structures, similar to nodes in the cosmic web, or halos. If the subspace $V(x)$ is the whole space and $\lambda_3>0$, we obtain underdense zero-dimensional structures similar to void centres. Finally, if the subspace $V(x)$ is spanned only by $v_3$ and $\lambda_3<0$, we obtain overdense two-dimensional structures similar to cosmic walls. It is important to note that these definitions always include the lower dimension objects: this means that filaments include nodes of the cosmic web (and therefore galaxy clusters) by definition.
        
        Alternative definitions of cosmic filaments can be found in the literature. In particular, it is common to classify the cosmic web according only to the eigenvalues of the hessian \citep{aragoncalvo2007,cautun2013, cautun2014}: all positive for halos, two positive for cosmic walls, one positive for filaments, and all negative for voids. Therefore, the space is completely divided into these four classes. This definition implies that regions classified as filaments are more clumpy, and are not one dimensional. In other words, it does not provide information about the centre of the filaments, which may be relevant for some applications. We note that ridges satisfy this definition, while also being local maxima in the direction perpendicular to the filament. Another common approach is to detect filaments with graph-based tools, such in \citet{bonnaire2019} and \citet{pereyra2019}. These methods do not define filaments using an underlying physical quantity such as density. Instead, they try to connect most galaxies with links subject to some regularising criteria. They result in a one-dimensional network, but they are usually sensitive to a range of parameters which must be tuned.

        \subsection{Tomographic analysis}
        \label{ss:tomo}
        
        The above definition of filaments can be applied to the cosmic web with $D=3$ dimensions by converting redshift into distance; or with $D=2$ dimensions by taking redshift slices and applying it independently to each slice. 
        
        Using a 3D galaxy distribution presents the problem of an asymmetric uncertainty, as the uncertainty in redshift is much larger than the uncertainty in sky position. This can generate false detections in the line-of-sight direction, especially near galaxy clusters \citep{kuchner2021}. It is also more computationally expensive in general, as the number of points that need to be stored in memory increases significantly.
        On the other hand, a tomographic approach is less sensitive to filaments with their axis aligned close to the line-of-sight direction. The angle between these directions will have an impact on the likeliness of a correct detection. Additionally, slicing the redshift range may produce border effects, such as the splitting of a long filament across two or more slices, or the detection of a filament in two consecutive slices when its redshift is close to the limit between slices.
        Depending on the goal, one method will be more suitable than the other.

        In this work, we use the second approach with thin slices in order to obtain tomographic information on the filament network. In principle, this means that we lose the ability to detect some of the filaments in exchange for a lower number of false detections, especially around clusters. Given that we work on spherical slices, all computations are done by taking into account the spherical geometry of the space, that is, the computation of the gradient, the hessian, and the definition of the density estimate $d$. 
        
        There are different techniques to slice the redshift information. An important aspect to take into account is that the width of the shells needs to be thin enough to avoid the case where different filaments overlap in a single slice; while it needs to be thick enough to include a large number of galaxies so we can detect the filaments in each slice. Additionally, we ensure that this width is larger than the redshift uncertainty of the galaxies within it by at least a factor of five.

        \subsection{Subspace-Constrained Mean Shift}
        \label{ss:scms}
        The SCMS algorithm exploits the previous definition of a ridge in order to find one-dimensional maxima. It has been used in cosmology to obtain cosmic filaments in galaxy catalogues \citep{chen2015a,chen2014} and simulations \citep{chen2015b}. This algorithm has also been applied to weak lensing maps by the Dark Energy Survey \citep{moews2020}; to galactic simulations in order to study filaments formed in interactions with satellite galaxies \citep{hendel2018}; and even in the analysis of the location of crimes in cities \citep{moews2019}.
        
        SCMS is an iterative algorithm. We start with a set of test points that uniformly fill the analysed area and at each step they move towards the filaments. In order to do that, we first compute an estimate of the galaxy density, and obtain its gradient and hessian. These points move in the direction given by the eigenvector of the hessian with the smallest eigenvalue. The magnitude of the movement is given by the projection of the gradient onto this eigenvector. In a ridge, the gradient is parallel to the direction of the ridge by definition, while this eigenvector is always perpendicular. This means that points remain fixed in place once they reach a ridge.
        
        A detailed description of the algorithm can be found in \citet{chen2015}. Here, we reproduce the steps as we have implemented them:
        \begin{alg}[filament detection]
                \label[alg]{alg:1}
                \begin{enumerate}
                        \item[\emph{In}:] A collection of points on the sphere, $\{x_i\} \equiv \{(\theta_i,\phi_i)\}$, representing the galaxies observed within a given redshift slice.
                        \item Obtain a density function $d(\theta,\phi)$ (defined over the sphere) by smoothing the point distribution with a certain kernel.
                        \item Calculate the gradient of the density: $g(\theta,\phi)\equiv\nabla d(\theta,\phi)$
                        \item Calculate the hessian of the density: $\mathcal{H} (\theta,\phi)$
                        \item Diagonalize the hessian ($2\times2$ matrix) at every point and find the eigenvectors $v_1(\theta,\phi)$, $v_2(\theta,\phi)$, with the smallest eigenvalue corresponding to $v_2$
                        \item Project the gradient $g$ onto the eigenvector $v_2$. Let $p(\theta,\phi)$ be this projection.
                        \item Select a set of points $\{y_j\}$ that will be used to search for the filaments. In our case, it will be a uniform grid over the sphere. Iterate until convergence:
                        \begin{enumerate}
                                \item At step $s$, move every point $y_j^{(s)}$ in the direction of the projection at that point $y_j^{(s+1)} = y_j^{(s)} + c \cdot p(y_j^{(s)})$. \footnote{The value of $c$ for optimal convergence varies with the pixel, and can be calculated with the derivative of the smoothing kernel. See \citet{comaniciu2002}.}
                        \end{enumerate}
                        \item When the points do not move between steps, they have reached a filament.
                        \item[\emph{Out}:] A collection of points on the sphere $y_j^{(S)}$ placed on the filaments.
                \end{enumerate}
        \end{alg}
        
\begin{figure*}
        	\centering
        	\includegraphics[width=0.32\textwidth]{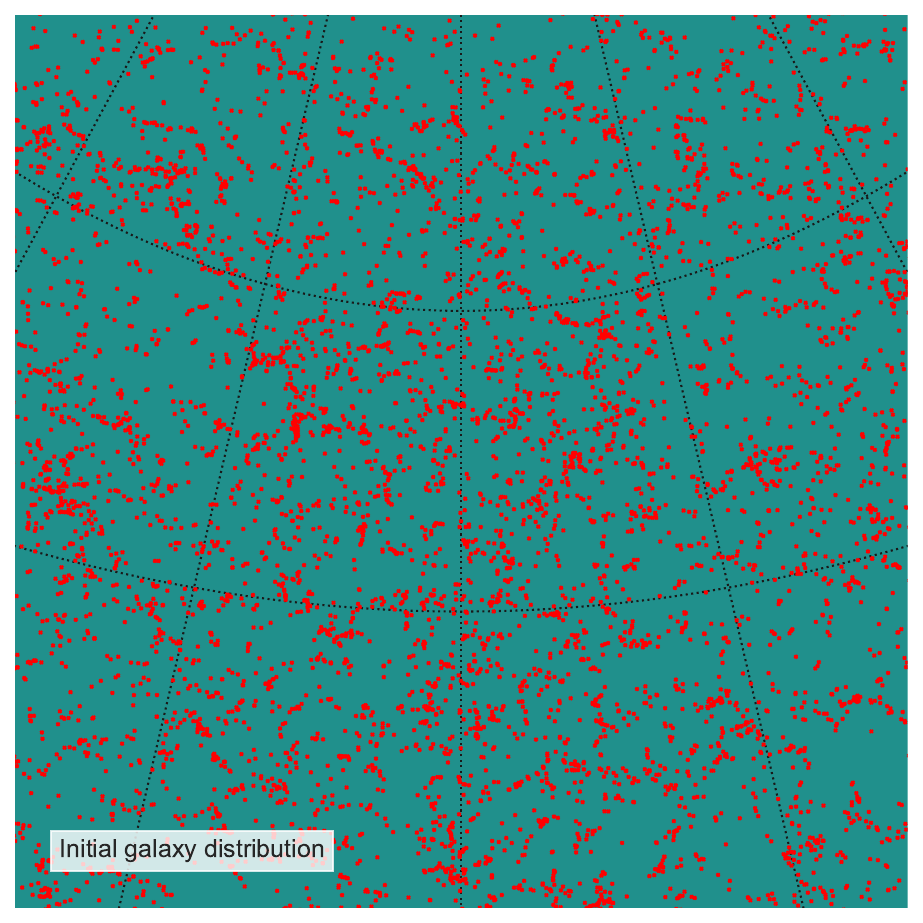}
        	\includegraphics[width=0.32\textwidth]{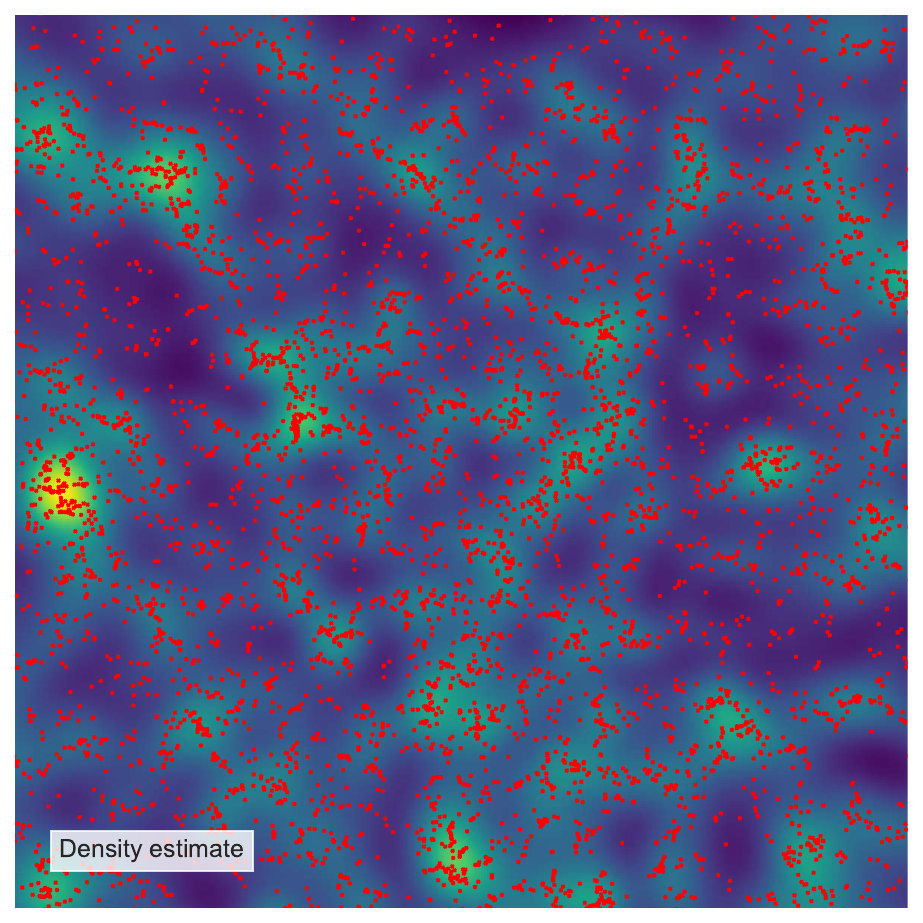}
        	\includegraphics[width=0.32\textwidth]{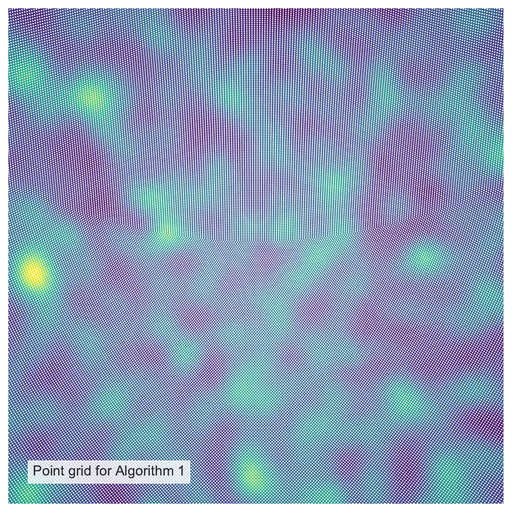}
        	
        	\includegraphics[width=0.32\textwidth]{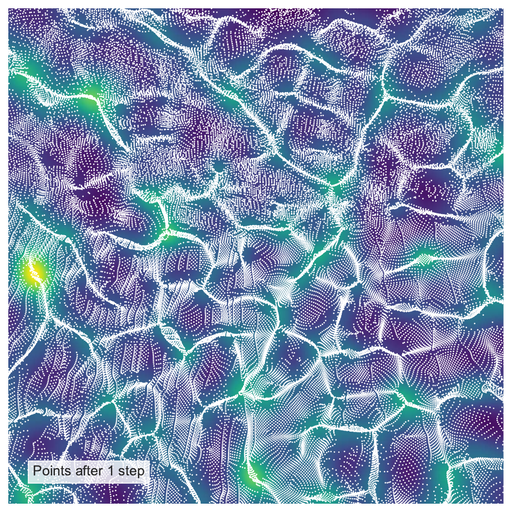}
        	\includegraphics[width=0.32\textwidth]{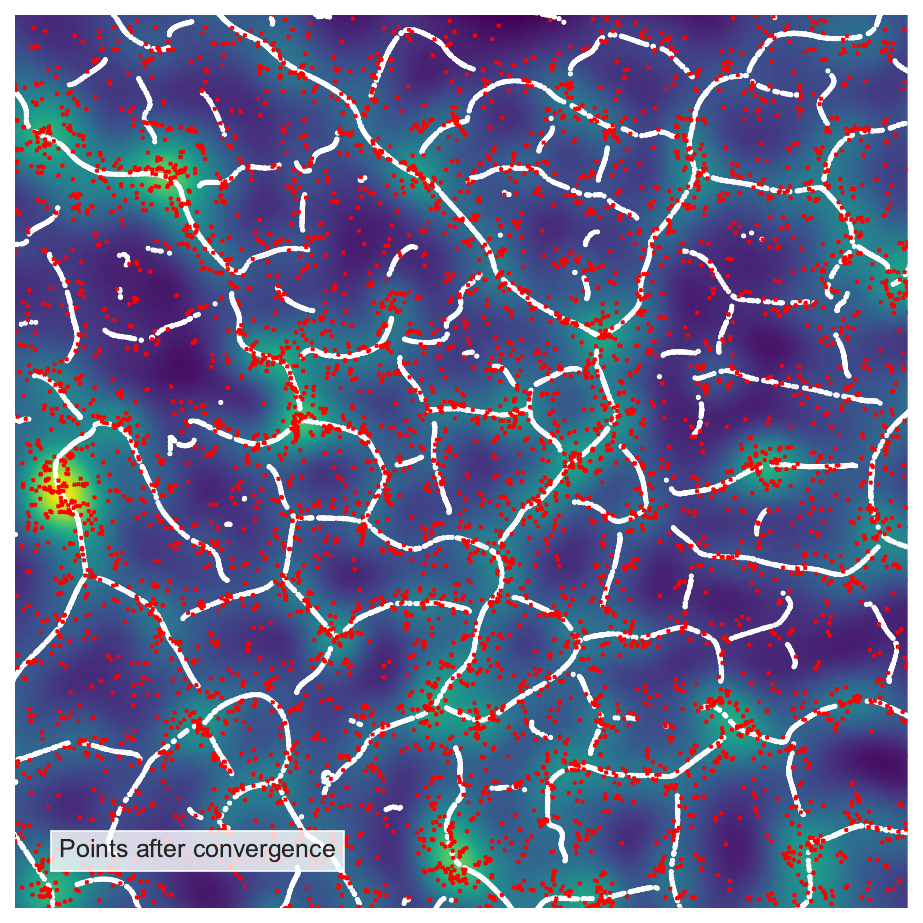}
        	\includegraphics[width=0.32\textwidth]{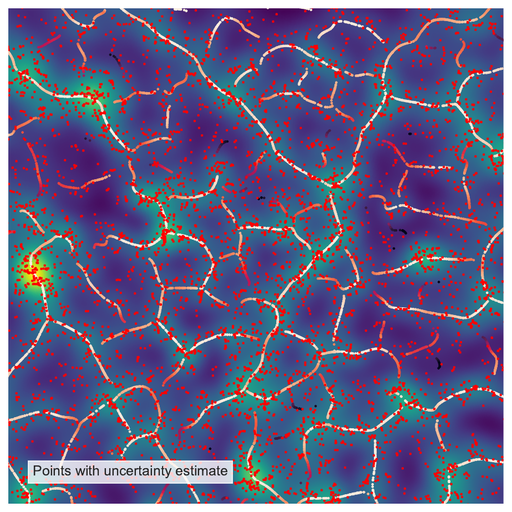}
        	\caption{Different steps of the algorithms. Top left: Initial distribution of galaxies in a redshift slice. Top centre: A density distribution is estimated from the galaxies. Top right: A grid of points $\{y_j\}$ is selected as the initial positions of the iterative steps; they correspond to the centres of HEALPix pixels. Bottom left: The positions of the points after one step of the iteration. Bottom centre: The positions of the points after they converge, with the initial galaxy distribution overlapped. Bottom right: Same as bottom centre, but the uncertainty of each point in the filaments is calculated with Algorithm 2; a redder colour represents a lower certainty. All images are gnomic projections centred at $RA=\ang{180}$, $dec=\ang{40}$ with a side size of \ang{60}. The data used in this example corresponds to BOSS data at $z=0.570$ (see \Cref{s:data}).
        		\label{f:example}}
        \end{figure*}

        \Cref{f:example} illustrates several steps of this algorithm.
        
All quantities defined over the sky ($d$, $g$, $\mathcal{H}$, $v_1$, $v_2$ and $p$) are computed using the HEALPix pixelisation scheme at resolution $N_{side}=1024$ \citep{gorski2005}. This corresponds to pixels with an approximate size of \SI{3.44}{\arcminute}. This is chosen to be much lower than the expected scales of any of the fields of interest (typically in the order of degrees) so as to avoid pixelisation effects.
        
        In order to implement the algorithm, we must choose the kernel to be used in the density estimation: in this work, we use a spherical Gaussian kernel (i.e. a Gaussian kernel evaluated on the geodesic distance). This requires the choice of a free parameter: the full width at half maximum (FWHM) of the kernel. The relation between the size of the kernel and the number of points can significantly affect the quality of the reconstruction in a redshift slice. To mitigate this effect, we implement two independent solutions: 
\begin{enumerate*}
\item[(a)] adapt the FWHM of the kernel to the number of points of each redshift slice; 
\item[(b)] combine the results of the algorithm at different scales using a machine learning approach. Both approaches are explained in \Cref{ss:size}. \end{enumerate*}

        \subsection{Estimate of the uncertainty}
        \label{ss:err}
        We measure the robustness of each detection as explained in \citet{chen2015}. This method is designed to simulate different realisations of the galaxy distribution using bootstrapping: for each simulation, we take a random sample of the real galaxies in order to obtain a modified realisation of galaxies.
        We then compare the filaments obtained in the real data with the filaments obtained in this new realisation of galaxies. In practice, we do this in the following way:
        
        \begin{alg}[uncertainty of the detection]
                \label[alg]{alg:2}
                \begin{enumerate}
                        \item The {true} filaments are computed on all the real data, $y_j$.
                        \item A new set of simulated galaxies is generated by bootstrapping the original galaxy catalogue. The new set has the same number of galaxies, but with possible repetitions. 
                        \item \Cref{alg:1} is run on these galaxies to obtain a {new} set of simulated filaments ${y^n_j}$.
                        \item For every point on the true filament, we compute the minimum distance to the closest new filament $\rho_n(y_j) = \min (d(y_j, y_j^n))$. Small distances correspond to more consistent detections.
                        \item We repeat steps 2 to 4 a total of $N=100$ times. For every point of the true filament, we have a minimum distance for each simulation; we find the (quadratic) mean of all simulations $\rho(y_j) = \sqrt{\frac{1}{N}\sum_{n=1}^{N} \rho_n(y_j)^2}$.
                        \item[\textit{Out}:] A single error estimate (average minimum distance) for every point in the true filaments, $\rho(y_j)$.
                \end{enumerate}
        \end{alg}
        
\begin{figure*}
                \centering
                \includegraphics[width=1\columnwidth]{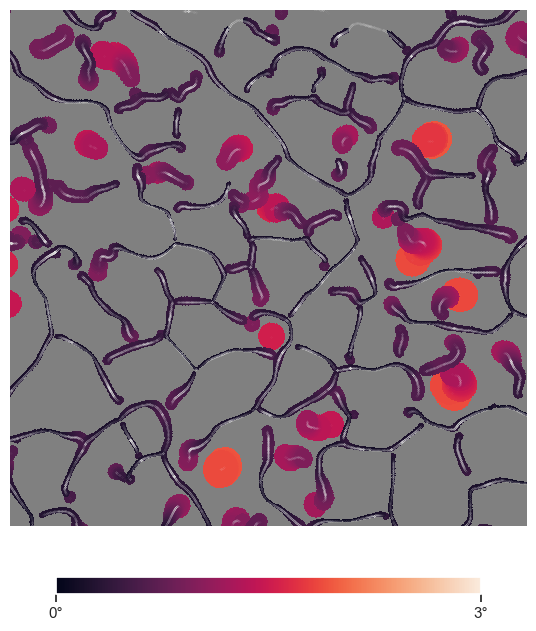} 
                \includegraphics[width=1\columnwidth]{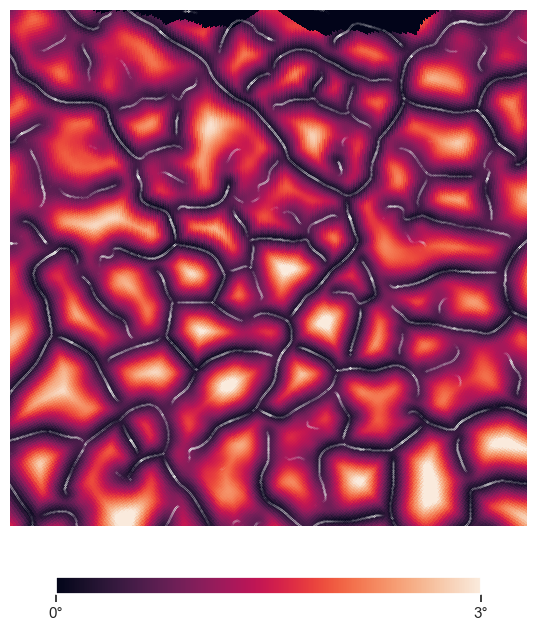} 
                \caption{Representations of the uncertainty of the detection. On the left, we plot a circle around each point of the filament with a radius equal to the uncertainty; this is a way of visualising the regions of the sky where a certain filament is expected to be. On the right, we apply the uncertainty estimate to each pixel of the map, and obtain an average distance to bootstrapped filaments for every pixel; actual filaments are found mostly where the mean distances are low. In both cases, colour represents uncertainty in degrees, and filaments found with real data are represented in white. Both figures are gnomic projections centred at $RA=\ang{180}$, $dec=\ang{40}$ with a side size of \ang{60}. The data used in this example correspond to BOSS data $z=0.570$ (see \Cref{s:data}).
                        \label{f:uncs}}
        \end{figure*}

        The value of $\rho(y_j)$ has a natural interpretation: it is the typical error in the location of a particular point in a filament, when the galaxy distribution changes slightly. As it is given in degrees, is can be readily represented as a confidence region on the sky. This is illustrated on the left image of \Cref{f:uncs}. We note that most filaments are determined with an accuracy well below \ang{1}, while the ends of filaments and smaller filaments present higher uncertainties. This uncertainty estimate may be very useful in applications where the user needs a certain position accuracy, as they can keep only the filaments that satisfy their criteria.
        
        We note that \Cref{alg:2} can be run on any set of points $y_j$, not necessarily the filament points. In this case, one obtains the typical distance to a filament for any point on the sky, averaged over the $100$ bootstrapping realisations. We use this approach to construct full maps of the regions of interest, setting $y_j$ to be the set of all pixels in the region of interest. This will be the basis of the second method to obtain filaments, as explained in the following section. An example of such a map can be seen in the right image of \Cref{f:uncs}. It can be seen that filaments are mostly detected in regions with minimum uncertainty. This figure also presents some spurious detections, which are characterised by high uncertainty and can be removed a posteriori.

        \subsection{Choice and combination of scales}
        \label{ss:size}
        
        As mentioned before, the shape and size of the filter kernel is a free parameter of \Cref{alg:1}. We use a Gaussian kernel to perform this step. We use two independent methods in order to select the best FWHM for each redshift slice, and mitigate the importance of this parameter; these will be explained in the following section.
        
        We note that we also tested the Mexican needlets of different orders \citep[see][]{narcowich2006,geller2008,marinucci2008}, as they have yielded better results than Gaussian profiles in several applications in cosmology \citep[e.g.,][]{oppizzi2019, planckcollaboration2016b}. However, in this particular case, they did not improve the results. This is due to the oscillating shape of these filters in pixel space, which may introduce artificial ridges in the maps.
        
        \subsubsection[Method A]{Method A: optimise the FWHM}
        \label{ss:size_A}
        
        The first method consists of finding an optimal value for the FWHM at each redshift slice. In order to avoid dependence on any theoretical model, we do not use the possible evolution of filaments as information to find them. The only parameter we consider is the number of galaxies per unit area within each redshift slice. Slices with a low density number do not have enough reliable information to reconstruct smaller filaments, and so the kernel has to be wider in order to compensate for this. On the other hand, highly populated slices contain enough information to reconstruct both large and small filaments, and so a narrower kernel can be used.
        
        The goal of changing the FWHM with redshift is to obtain the most complete reconstruction possible at each redshift, while maintaining a similar quality of the reconstruction in terms of accuracy and error.
        
        In practice, we scale the FWHM with the number of galaxies using the following expression:
        \begin{equation}
                fwhm = A \cdot \left(\frac{n}{f}\right)^{-\frac{1}{6}}
        ,\end{equation}
        where $n$ is the number of galaxies within a redshift slice, $f$ is the observed sky fraction, and $A$ is a constant parameter. The exponent $-1/6$ is a common scaling factor for kernel density estimators in two dimensions, and is used for example in the \textit{Silverman's rule of thumb} \citep{silverman1998}. This is proven to be optimal in the case of reconstruction of a Gaussian profile, in which case $A$ can be optimally calculated. However, as we do not expect a Gaussian profile, we tune the prefactor $A$ of this rule of thumb to
        \begin{equation}
                A  = 21\, \deg .
        \end{equation}
        We tested different values in simulations with filaments and noise. This value yields better results than higher ($28, 35$) or lower ($14$, $7$) values; with a false detection rate below $15\%$ and discovery power above $90\%$. This value seems to perform better with real data as well.
We refer to the filament catalogue obtained with method A as Catalogue A.

        \subsubsection[Method B]{Method B: combine the information at different scales}
        \label{ss:combine_B}
        
        The alternative method aims to eliminate the choice of a FWHM by combining different scales. It combines the results of filtering at different scales in order to obtain a single predictor of the filament locations. The key idea is that the quantities computed in \Cref{alg:1} at different scales (density, gradient, hessian, etc.) contain enough information in order to be able to distinguish a point in a filament from a point far from it. By combining these quantities, we may be able to predict the location of filaments more accurately than applying the algorithm on a single scale. Multiscale approaches have been used before in the literature \citep{aragoncalvo2007}; in this work we develop a powerful new method to combine the information into a single output.
        
        In order to combine the information at different scales, we deploy a machine learning algorithm: Gradient Boosting. This algorithm generates a series of \textit{decision trees} based on the input values for each quantity. The first tree will be trained to predict the desired output, the second tree is trained to learn the error caused by the first, the $n$-th tree is trained to predict the error of the sum of the first $n-1$. By generating a large number of these trees and combining their predictions in this way, the ensemble prediction is greatly improved. For more information about the algorithm, see \citet{tkh1995}, \citet{friedman2001}, and \citet{geron2019}. The choices for some of the most important hyperparameters are the following: number of trees ($100$), learning rate ($0.1$), loss function (least squares), and maximum depth of each tree ($3$ layers). These are the baseline settings for this algorithm and we have verified results are stable when faced with small changes in these parameters. This would resolve the problem of fixing the FWHM and eliminate the dependence with redshift. Additionally, the algorithm should be able to learn the typical characteristics of filaments in the most populated redshift slices at low redshift, where filaments are easier to detect. It can then use this learnt information in order to identify filaments at higher redshift or less populated slices. Galaxies at lower redshift (red galaxies) and higher redshift (quasars) are tracers with different characteristics: quasars have a higher value of the bias \citep{laurent2017}. In order to train and apply the algorithm on the whole redshift range, we must ensure that the algorithm is flexible enough to accommodate both datasets. We confirm that this is the case in \Cref{ap:lrg-qso}.

        The idea of this algorithm is similar to the well-established technique of convolutional neural networks (see \citet{krachmalnicoff2019} for an implementation of convolutional neural networks compatible with the HEALPix pixelisation). These networks typically convolve the map with a series of stacked layers; each layer containing a number of small filters trained to maximise the information extracted from the map. In order to extract information from a large area around a pixel, a large number of layers must be stacked. However, in our case, we can extract more useful information from the density map by obtaining relevant quantities such as the modulus of the gradient or the eigenvalues of the hessian at different scales, as these are very good indicators of the location of ridges. These quantities cannot be easily reproduced by training linear filters. Computing these quantities also implies isotropic filters, as opposed to trained filters, where the orientation is important. Lastly, extracting these features explicitly allows us to use different fixed scales; these scales are much larger than the pixel size ($\sim50$ to $100$ times larger), which can usually  only be obtained by stacking a large number of convolutional layers.

        The input of the Gradient Boosting is some of the quantities computed in \Cref{alg:1} at each pixel. We select five quantities that could be physically correlated (positively or negatively) with the presence of a filament: the density estimate $d$, the magnitude of the gradient $\left\| g \right\| $, the magnitude of the projection $\left\| p \right\| $, the value of the second eigenvalue of the Hessian of the density $\lambda_2$, and its \textit{eigengap} $\lambda_1-\lambda_2$. We compute these quantities for each pixel at several smoothing scales. For example, for the data we analyse here, we verified that four scales provide a stable reconstruction, while adding more scales does not carry any significant improvement. In particular, we use FWHM $= $\;\SIlist{2.7; 3.4; 4.1; 4.8}{\deg}. Therefore, there are $20$ input features ($5$ quantities times $4$ scales) and as many points as pixels in the region of interest.
        
        As output of the Gradient Boosting, we want a measure of the `filament-ness' for each pixel. In order to do this, we exploit the method to estimate the error described in the last paragraph of \Cref{ss:err}. There we derived a map in which the value of each pixel is the average distance to a filament in the bootstrapping simulations, as in the right image of \Cref{f:uncs}. The machine learning algorithm uses the value of the quantities at each pixel to predict the typical distance of a filament from a pixel with certain characteristics. In particular, the algorithm learns the characteristics of the pixels with filaments. The exact procedure to produce the training maps is explained in \Cref{ap:train}.

        After the algorithm is trained, we obtain a prediction of the estimated error map for each redshift slice. In order to compare with the previous method, we obtain a filament distribution by applying \Cref{alg:1} again on these maps. We report the filament catalogue, which we call Catalogue B.
        
        There are three
key advantages to making the prediction with the Gradient Boosting algorithm:
        \begin{itemize}
                \item It generalises better. The algorithm is able to learn what range of values are expected at a filament, close to a filament, and far from a filament. This means that results from redshift slices with a greater number of observed galaxies will improve the algorithm at all redshift slices. In particular, the algorithm becomes more robust at higher redshifts where there are fewer observations.
                \item It reduces the impact of anomalies. For the same reason, spurious detections become less likely, as they are sometimes caused by local ridges that do not correspond to values expected from real filaments (e.g. produced by a small number of galaxies).
                \item It is much faster than computing the real scale-combined estimated uncertainty maps, as explained in \Cref{ap:train}.
        \end{itemize}

        \subsection{Strengths and limitations}
        \label{ss:pros} 
        
        Given the choices explained in this section and the algorithms used, we identify the following strengths ($+$) and limitations ($-$) of our methodology:
        \begin{itemize}
                \item[$+$] The implementation works natively in the surface of the sphere, taking into account its geometry, and is compatible with the widely used HEALPix scheme. This also makes the implementation rotationally invariant and independent of the chosen reference system.
                \item[$+$] We reduce the sensitivity to the density estimate thanks to the development of a boosted version of the standard SCMS algorithm with machine learning (Method B).
                \item[$+$] The approach provides a tomographic catalogue. This eases the study of the evolution of different parameters or characteristics of filaments as a function of redshift. Additionally, it avoids or reduces artefacts due to the conversion from redshift to distance, such as the \textit{Finger of God} effect or the asymmetrical error in the 3D\ location.
                \item[$+$] The method provides uncertainty estimates for each detection using bootstrapping simulations (see \Cref{ss:err}).
                \item[$+$] The algorithm is fast, it takes only a few seconds per redshift slice on a laptop.
                \item[$\pm$] Physical centre and width: the ridge definition applies to the centre of the filaments, and so the algorithm is able to locate the 1D centre; however, it gives no direct information about its physical width.
                \item[$-$] This methodology is most sensitive to filaments perpendicular to the line of sight direction. Filaments with directions close to the line of sight will not be detected by this methodology.
                \item[$-$] The ridge definition does not determine the individual filaments, it only determines which points are in a filament; filament separation needs to be done afterwards using another method.
                
        \end{itemize}

        \section{Data}
        \label{s:data}

        \begin{figure}
                \centering
                \includegraphics[width=0.9\columnwidth]{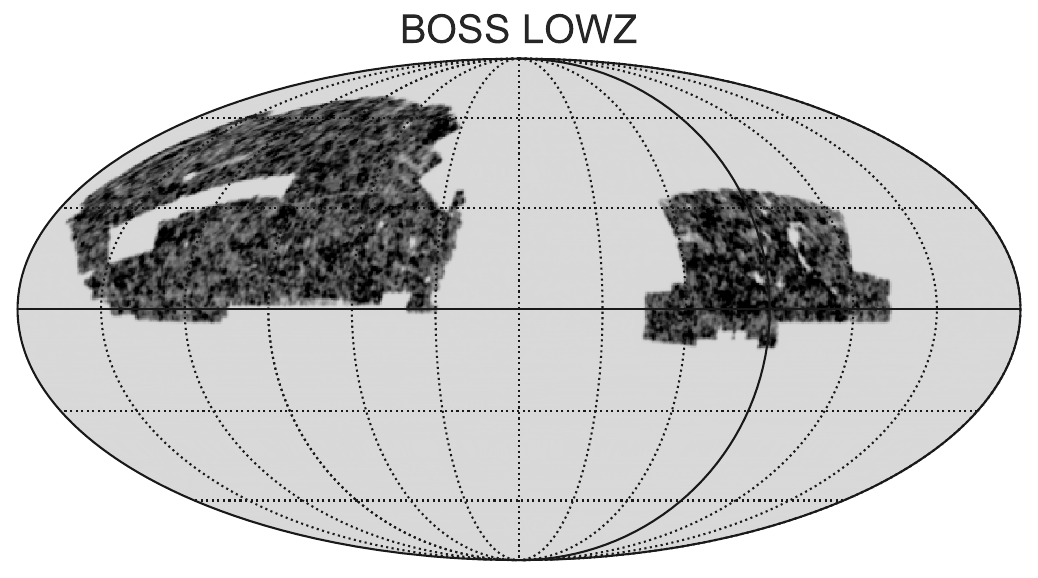} 
                \includegraphics[width=0.9\columnwidth]{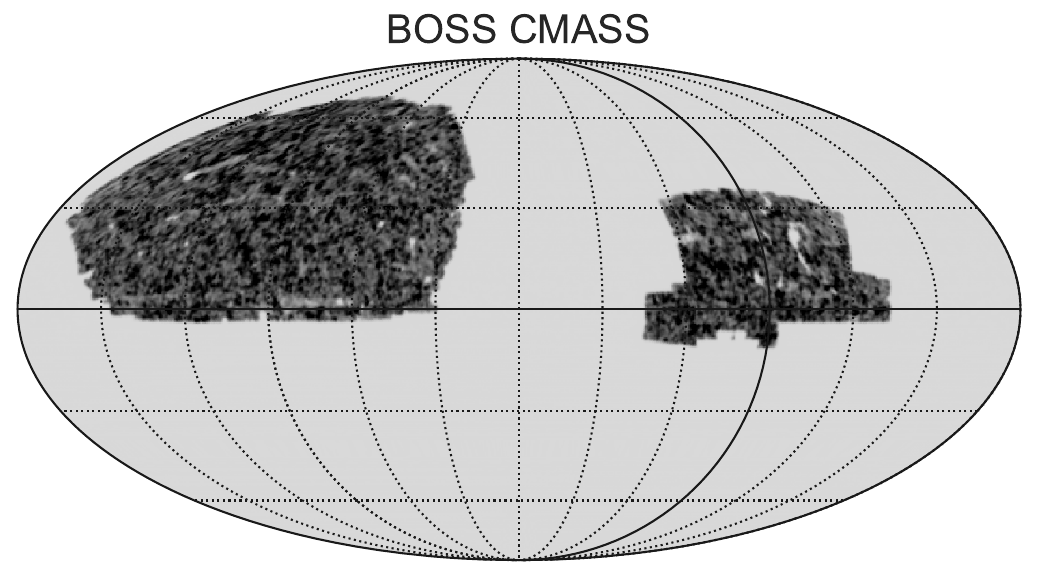} 
                \includegraphics[width=0.9\columnwidth]{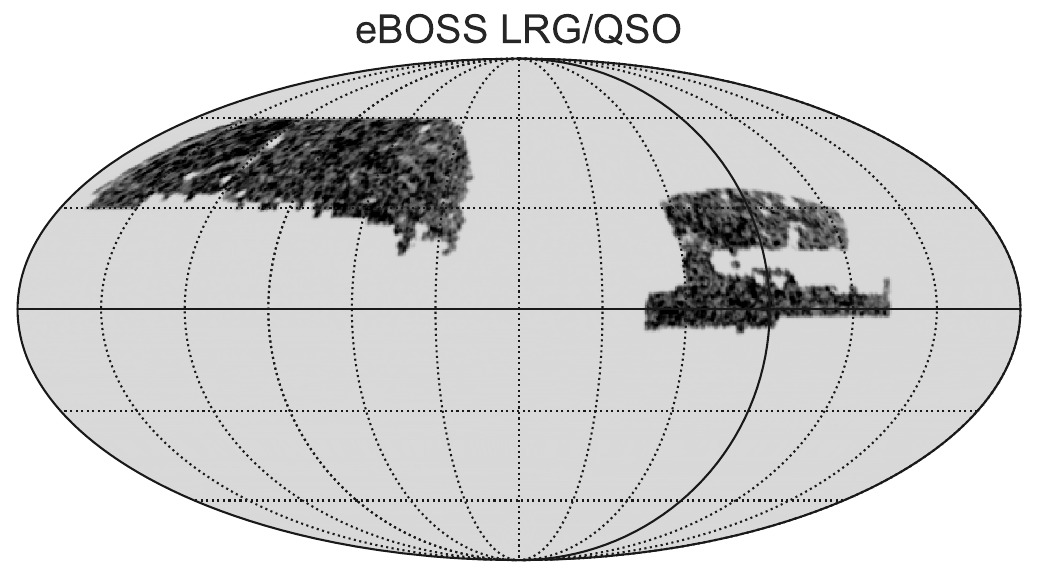} 
                \includegraphics[width=0.9\columnwidth]{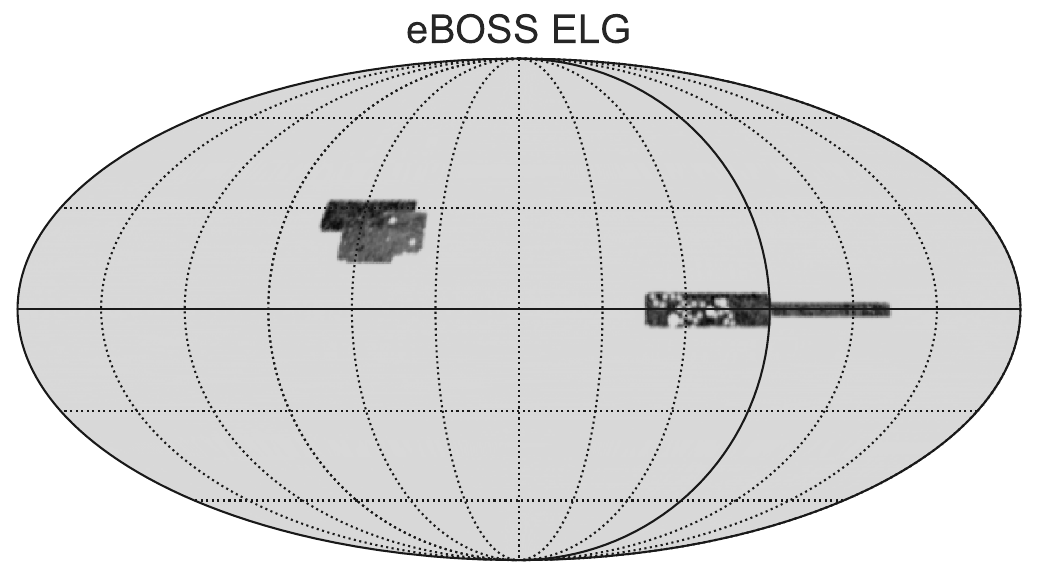} 
                \caption{Sky footprint of the data reported by BOSS LOWZ (top), BOSS CMASS (second), and the three eBOSS samples: LRG and QSO (third), and ELG (bottom). The representation uses a mollweide projection in equatorial coordinates (J2000), with a rotation of \ang{90} along the $z$ axis. The footprints on the left and right sides correspond to the north and south Galactic caps, respectively.
                        \label{f:skies}}
        \end{figure}

        The Sloan Digital Sky Survey (SDSS, \citealp{blanton2017}) is currently the largest and most complete spectroscopic galaxy catalogue. In particular, we use two of its surveys: BOSS (Baryon Oscillation Spectroscopic Survey) and eBOSS (extended BOSS). Both of these surveys report a specific catalogue tailored to the study of large-scale structure. They are divided into different samples, each optimised to a specific redshift range. In this section we briefly describe the different samples and how we use and combine them.
        
        \subsection{BOSS}
        
        BOSS maps the distribution of luminous red galaxies (LRG) and quasars (QSO) on the sky and in redshift. The final data release is part of the SDSS DR12 \citep[see][]{dawson2013,alam2015,anderson2014}. They report two samples:
        \begin{itemize}
                \item LOWZ: Galaxies in this sample are mostly located in the range of $0.05<z<0.5$ in both the north and south Galactic cap. The sky footprint can be seen in \Cref{f:skies} (top row). The sky fraction is approximately $17.0\%$ and $7.5\%$ for the north and south Galactic caps, respectively. The north Galactic cap presents an unobserved region; we note that we remove all filaments detected close to this region in order  to avoid border effects, see \Cref{ap:clean}.
                \item CMASS: Galaxies in this sample are mostly located in the range of $0.4<z<0.8$ in both the north and south Galactic caps. The sky footprint can be seen in \Cref{f:skies} (second row). The sky fraction is approximately $19.0\%$ and $7.5\%$ for the north and south Galactic caps, respectively.
        \end{itemize}
        
        \subsection{eBOSS}
        The eBOSS survey uses three different tracers: LRGs, emission
line galaxies (ELGs), and QSOs. The most recent data are reported in SDSS DR16 \citep[see][]{dawson2016,ahumada2020,ross2020}. Catalogues for each type are reported separately:
        \begin{itemize}
                \item LRGs: Galaxies in this sample are located in
the range of $0.6<z<1.0$ in both the north and south Galactic cap. The sky footprint can be seen in \Cref{f:skies} (third row). The sky fraction is approximately $8.4\%$ and $6.0\%$ for the north and south Galactic caps, respectively. The north Galactic cap is a subregion of the BOSS footprint; it is around half the size of this latter and is very regular. The south Galactic cap is irregular and presents a gap. Given that the area close to borders may introduce spurious detections (\Cref{ap:clean}), the region available for the analysis is too small to be used for our filament-finding algorithm: we do not include this cap.
                \item ELGs: Galaxies in this sample are located in
the range of $0.6<z<1.1$ in both the north and south Galactic cap. The sky footprint can be seen in \Cref{f:skies} (bottom row). The sky fraction is $1.6\%$ and $2.0\%$ for the north and south Galactic caps, respectively. Both of these caps are too small and irregular for the algorithm, so we do not use this sample.
                \item QSOs: Quasars in this sample are located in
the range of $0.8<z<2.2$ in the north and south Galactic cap. The sky footprint is identical to the one for LRG, which can be seen in \Cref{f:skies} (third row). The sky fraction is approximately $8.4\%$ and $6.0\%$ for the north and south Galactic caps, respectively. As before, we are unable to use the data in the south Galactic cap, and so we  use only the north Galactic cap.
        \end{itemize}

        \subsection{Combining data}
        \label{ss:combine}
        In order to produce the best possible filament catalogue, we carefully combine the previous data, maximising the density of galaxies and the covered sky fraction. Combining different samples by simply merging the catalogues may generate an inhomogeneous sample on the sky. This would introduce artefacts in the regions when the number of galaxies varies abruptly due to sampling inhomogeneities. Therefore, when combining samples with different footprints, we need to select the most restrictive footprint. This means that there is a natural trade-off between the observed number of galaxies and the sky footprint to be considered.

        \begin{figure*}
                \centering
                \includegraphics[width=\textwidth]{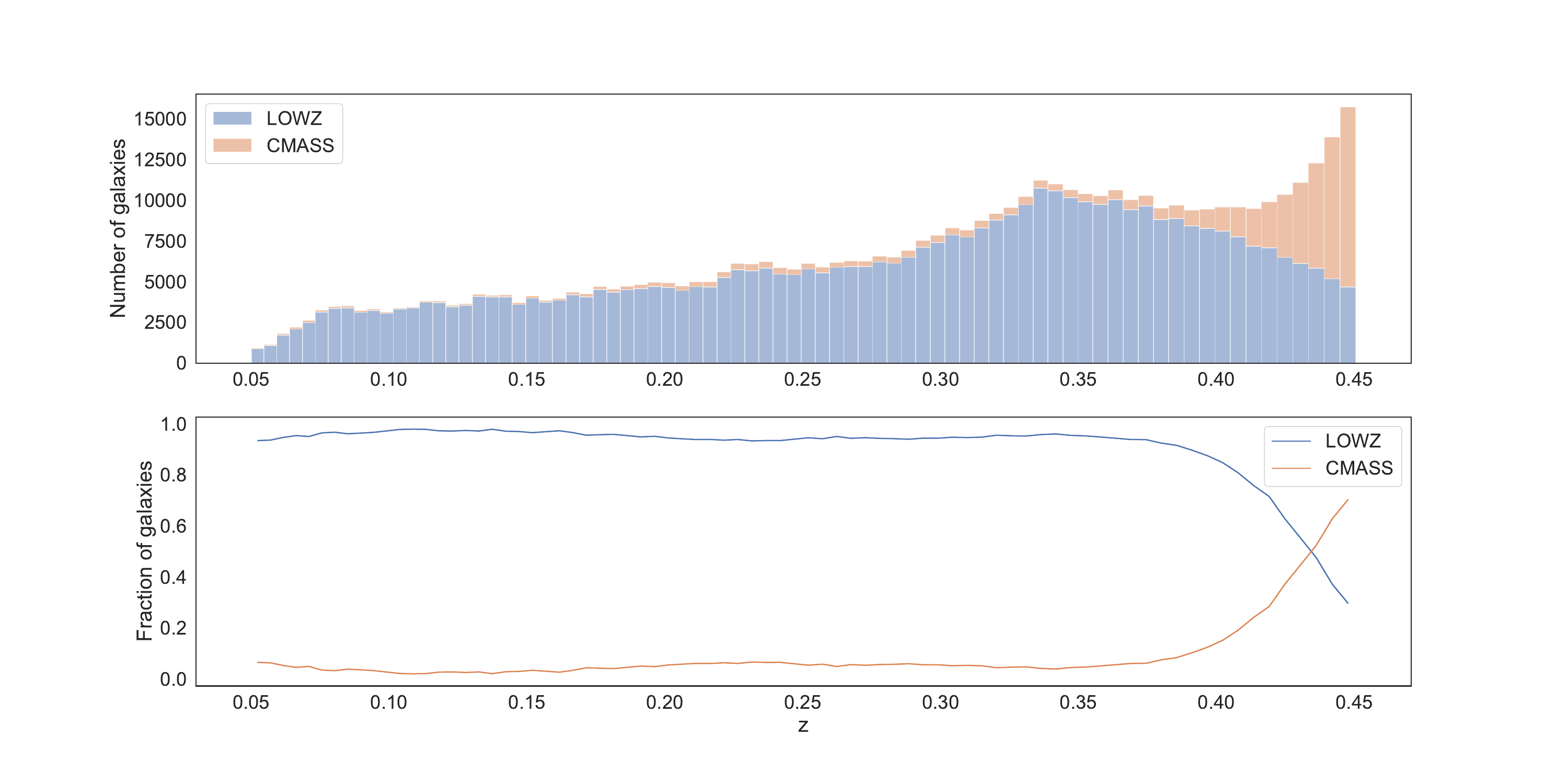}
                \caption{Galaxy distribution in Block 1. Top: Stacked number of galaxies per redshift bin from both the used surveys. Bottom: Fraction of galaxies corresponding to each survey as a function of redshift.
                        \label{f:block1}}
        \end{figure*}
        
        \begin{figure*}
                \centering
                \includegraphics[width=\textwidth]{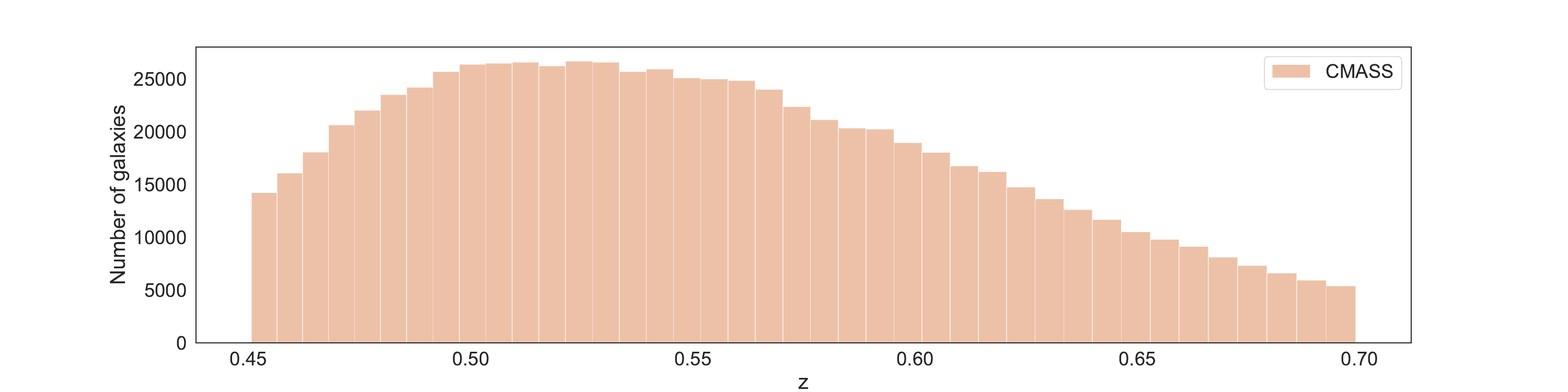}
                \caption{Galaxy distribution in Block 2. Number of galaxies per redshift bin. All galaxies correspond to the CMASS sample.
                        \label{f:block2}}
        \end{figure*}

        \begin{figure*}
                \centering
                \includegraphics[width=\textwidth]{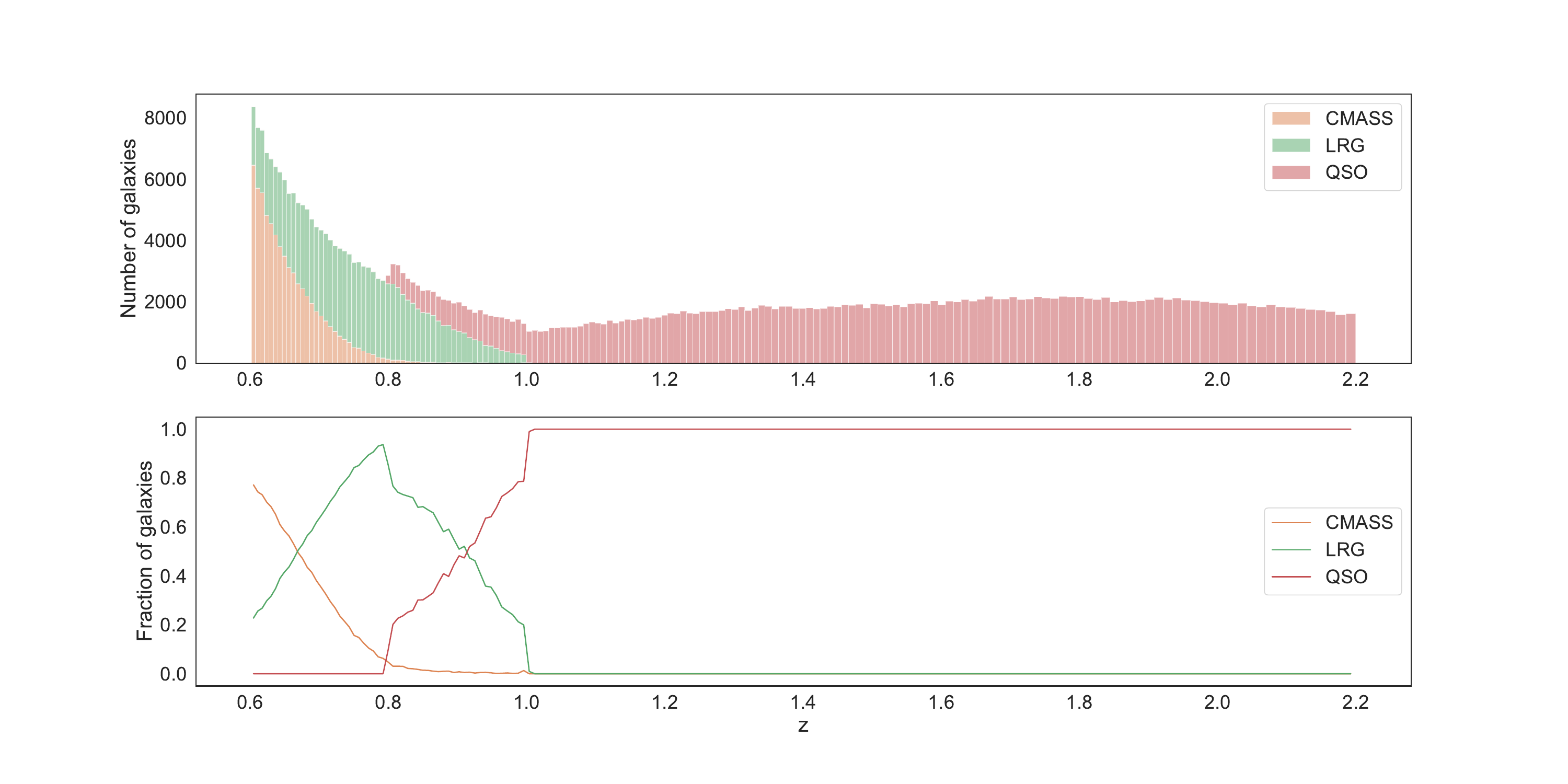}
                \caption{Galaxy distribution in Block 3. Top: Stacked number of galaxies per redshift bin from all of the used surveys. Bottom: Fraction of galaxies corresponding to each sample as a function of redshift.
                        \label{f:block3}}
        \end{figure*}

        In practice, we combine the data into the following Blocks (i.e. data samples), each with a different mask:
        \begin{itemize}
                \item Block 1: $0.05<z<0.45$. Data from LOWZ and CMASS; in the sky footprint from LOWZ, which is the most restrictive of the two. The number of galaxies per redshift bin and the ratio of both samples can be seen in \Cref{f:block1}. We note that this block is dominated by LOWZ, with CMASS data providing a relevant contribution (more than $10\%$ of the galaxies) only at $z>0.4$. At the higher redshift end, the majority of galaxies come from the CMASS sample. However, there are a large number of LOWZ galaxies which we decide not to neglect as they improve the filament reconstruction, although this does reduce the sky region slightly.
                \item Block 2: $0.45<z<0.7$. Data from CMASS only. The number of galaxies per redshift bin can be seen in \Cref{f:block2}.
                \item Block 3: $0.6<z<2.2$. Data from CMASS, LRG and QSO; in the footprint from LRG (equal to the QSO footprint), the most restrictive of the two regions. There are no QSO data below $z=0.8$, while all filaments above $z=1.0$ are detected using QSO data exclusively. The number of galaxies per redshift bin and the fraction of each sample can be seen in \Cref{f:block3}.
        \end{itemize}

        We note that the redshift range $0.6<z<0.7$ is covered by two Blocks: Block 2 which is made up only by CMASS data in a larger sky region; and Block 3 which is made up by CMASS+LRG data, but in a smaller region given by LRG. We perform the filament reconstruction in this redshift range separately for the two Blocks. This is done in order to extract the most information out of the samples, while ensuring that the entire region is homogeneously sampled in order to avoid the introduction of artefacts. As an interesting byproduct, we are able to compare filaments extracted from BOSS alone and from BOSS+eBOSS.

        \section{The filament catalogue}
        \label{s:results}
        \subsection{Filament reconstruction}
        We apply Methods A and B, as explained in \Cref{ss:size}, to extract the cosmic filaments catalogues from the three Blocks of data introduced in \Cref{ss:combine}. These three Blocks correspond to three different redshift ranges and sky fractions $f$ inherited from the used SDSS data: (a) Block 1 ($0.05<z<0.45$), with $78$ bins and $f_1=0.186$; (b) Block 2 ($0.45<z<0.7$), with $40$ bins and $f_2=0.219$; and (c) Block 3 ($0.6<z<2.2$), with $166$ bins and $f_3=0.062$. 
        
        The width of the slices in the line-of-sight direction is taken as $20$ Mpc, computed with the $\Lambda$CDM parameters obtained by Planck \citep{planck2019vi}; in particular, $H_0=67 \frac{km}{s \, Mpc}$. We note that the exact parameters to be used have a limited impact, as small variations of this width do not significantly alter our results. This width is chosen to be small enough to avoid several overlapping filaments in a single slice, but large enough to be several times larger than the uncertainty on the galaxy measured redshift.
        
\begin{figure*}
                \centering
                \includegraphics[width=0.9\textwidth]{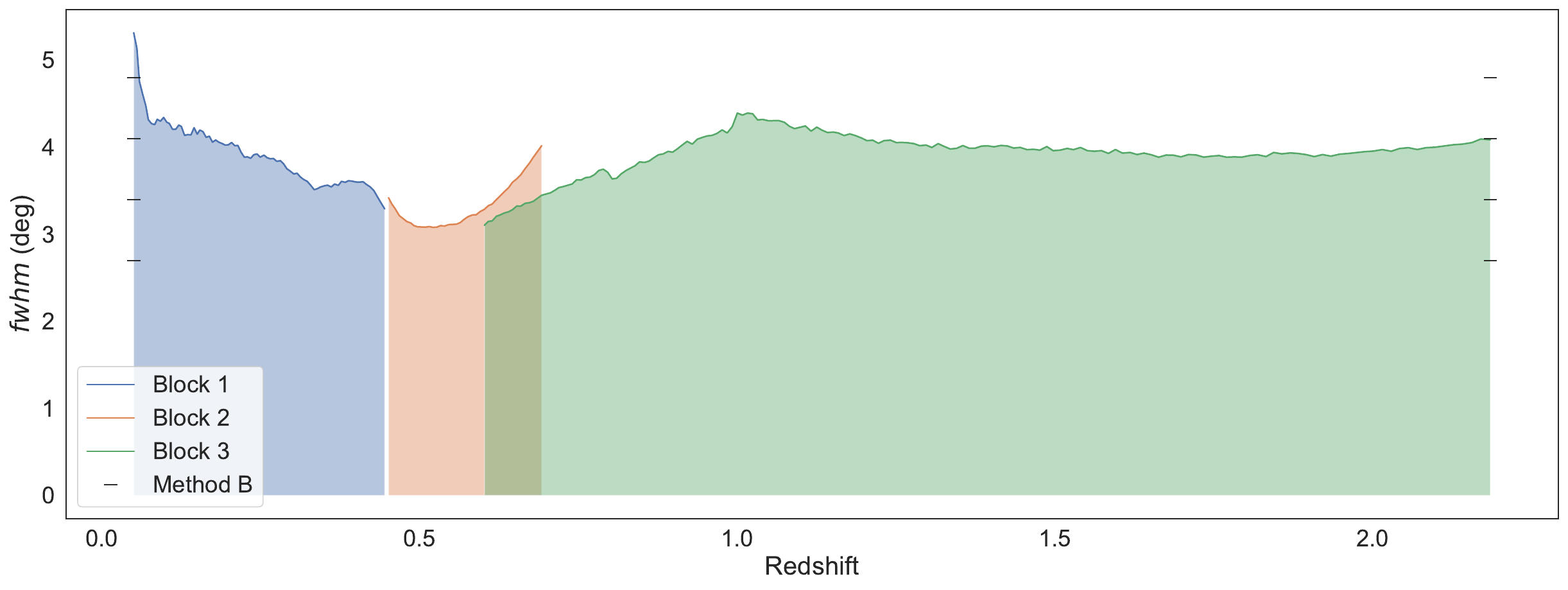}
                \caption{Value of the FWHM of the Gaussian kernel chosen for Method A as a function of redshift for each block. The scales that are combined in Method B are fixed and can be seen with markers on both redshift ends.
                        \label{f:fwhm}}
        \end{figure*}

        Method A requires the choice of the size of the smoothing kernel. As explained in \Cref{ss:size_A}, this is taken to be a function of the number of galaxies within a redshift slice. The value of FWHM of the Gaussian kernel, in degrees, can be seen in \Cref{f:fwhm}. Method B uses four fixed scales which can also be seen in the figure ($2.7\, \deg, 3.4\, \deg, 4.1\, \deg, 4.8 \, \deg$); they contain the range of sizes used for Method A.

        We compare the results obtained with both methods in \Cref{ss:comp}. We see that Method B generally outperforms Method A. Therefore, we recommend the use of the catalogue produced with Method B, which is publicly available at \href{https://www.javiercarron.com/catalogue}{javiercarron.com/catalogue}. The catalogue produced with Method A is also available upon request.
The columns of the catalogues are explained in \Cref{t:cata}.

        \begin{table}
                \caption{Columns in the catalogues.
                        \label{t:cata}}
                \centering
                \begin{tabular}{ll}
                        \textbf{Column} & \textbf{Name}  \\
                        RA & Right ascension ($\deg$) \\
                        dec & Declination ($\deg$) \\
                        dens & Density estimate  \\
                        unc & Estimated uncertainty ($\deg$) \\
                        grad\_RA & Gradient (RA component)  \\
                        grad\_dec & Gradient (dec component)  \\
                        angle & Angle of the filament ($\deg$) \\
                        z\_low & Start of the redshift bin  \\
                        z\_high & End of the redshift bin  \\
                        ini\_dens & Initial density estimate \\
                \end{tabular}
            \tablefoot{\textit{RA} and \textit{dec} refer to the equatorial system J2000. \textit{Angle} refers to the angle of the filament with the parallels of the sphere in these coordinates, in a counter-clockwise direction. Column \textit{dens} refers to the density estimate of the filament. Column \textit{ini\_dens} refers to the density estimate of the starting points at the beginning of \Cref{alg:1}, before the points move towards the ridges; this can be used to set a threshold, see \Cref{ap:clean}. Units of density estimates are arbitrary.}
        \end{table}

        \subsection{Comparison between methods}
        \label{ss:comp}
        In this section we present and compare the catalogues obtained with Methods A and B explained in \Cref{ss:size}. We use the final results in the complete redshift range $0.05<z<2.2$ and all available sky.
        
\begin{figure*}
                \centering
                \begin{subfigure}{0.45\textwidth}
                        \centering
                        \includegraphics[width=\textwidth]{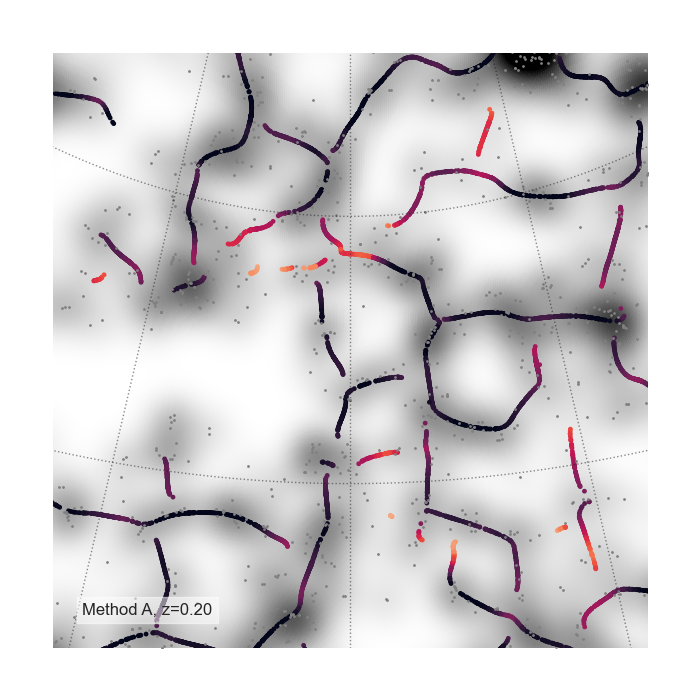}
                \end{subfigure}
                \begin{subfigure}{0.08\textwidth}
                        \centering
                        \includegraphics[width=\textwidth]{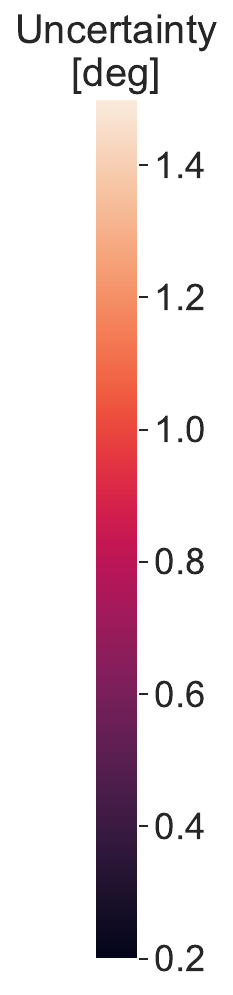}
                \end{subfigure}
                \begin{subfigure}{0.45\textwidth}
                        \centering
                        \includegraphics[width=\textwidth]{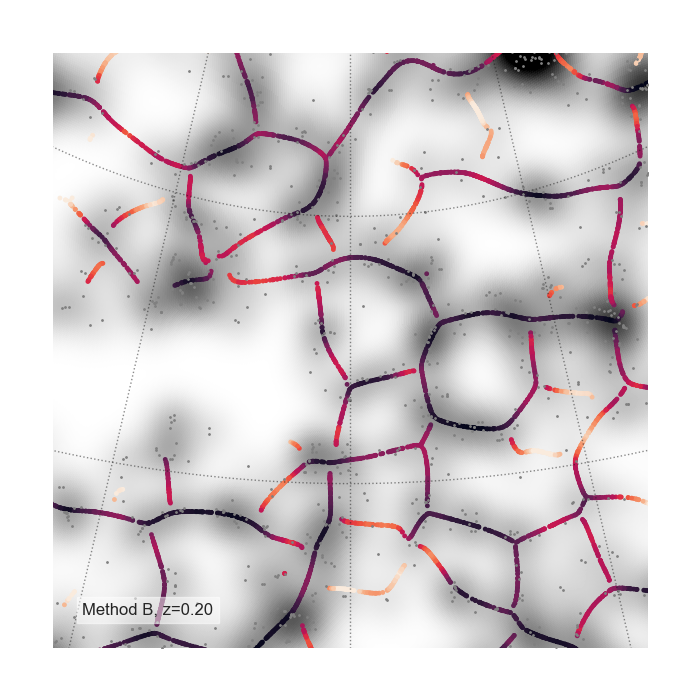}
                \end{subfigure}
                
                \begin{subfigure}{0.45\textwidth}
                        \centering
                        \includegraphics[width=\textwidth]{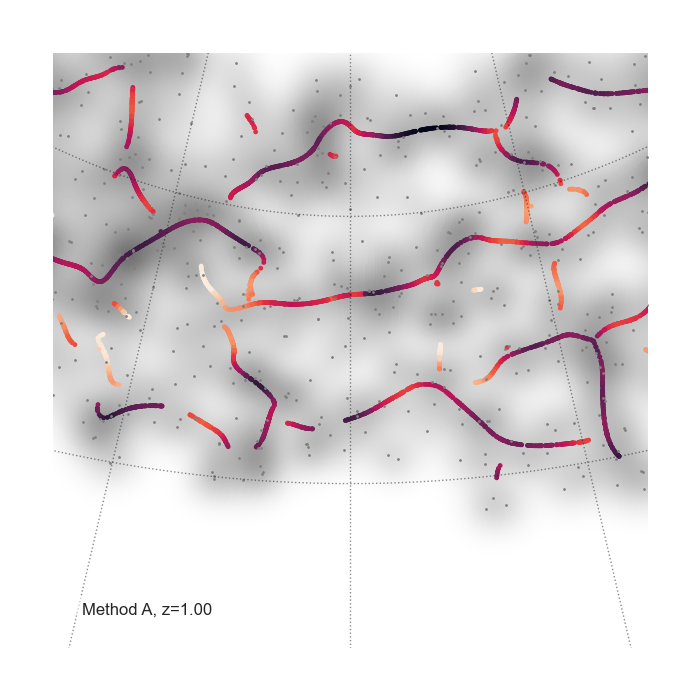}
                \end{subfigure}
                \begin{subfigure}{0.08\textwidth}
                        \centering
                        \includegraphics[width=\textwidth]{color}
                \end{subfigure}
                \begin{subfigure}{0.45\textwidth}
                        \centering
                        \includegraphics[width=\textwidth]{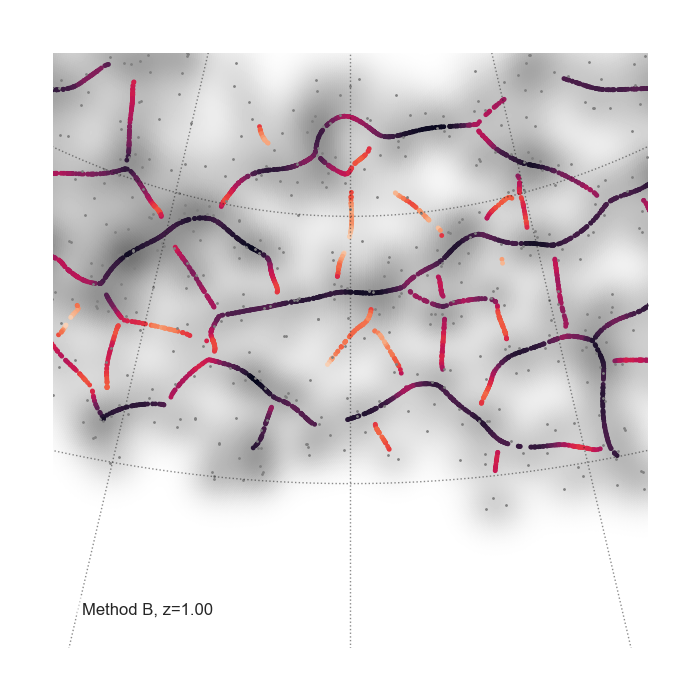}
                \end{subfigure}
                
                \caption{Example of reconstructed filaments at $z=0.20$ (top) and $z=1.00$ (bottom); with Method A (left) and Method B (right).
                        The colour of the points represents the error estimate in degrees. Grey points correspond to galaxies, while the background map represents the galaxy density, smoothed at \ang{3}, for visualisation purposes. The visualisation is a gnomonic projection around the point ($RA=\ang{140}$, $\delta=\ang{40}$) with a side size of \ang{60}. Meridians and parallels are both represented at a step of \SI{20}{\deg}.
                        \label{f:comp}}
        \end{figure*}

        The first difference between the two catalogues is that Method B is able to detect more filaments. An example at $z=0.20$ can be seen in the top row of \Cref{f:comp}; this is representative of the results at lower redshift ($z<0.5$). It can be seen that most filaments in Catalogue A are also detected in Catalogue B. The latter have a slightly larger associated uncertainty. However, more importantly, there is an additional filament population in Catalogue B, which is detected at lower significance. These additional filaments are not detected with Method A because they are erased when smoothing the galaxy density map at a single scale. On the other hand, as expected, Method B is able to correctly recover the same filaments detected with method A, plus additional filaments from other (usually smaller) scales. 
        
        At higher redshift ($z>0.7$), we continue to observe an additional filament population for the the same reason. However, the filaments which are observed in both catalogues are detected with a significantly lower uncertainty estimate in Catalogue B. A representative example at $z=1.00$ can be seen in the bottom row of \Cref{f:comp}. At this redshift, the number of galaxies is lower; Method A only uses the information at the given slice, and so it is more sensitive to noise and incomplete information. On the other hand, the machine learning algorithm in Method B is able to `learn' the typical characteristics of filaments in slices where there are more galaxies and less noise. It is then able to use some of this information in all redshift slices to obtain a more confident and uniform reconstruction of the filaments.
        
        \subsubsection{Distance from galaxies to filaments}
        \label{ss:distance}
        The filament catalogues are built to trace the regions with a galaxy number overdensity. Consequently, there is a correlation between filaments and galaxies by construction. We use this correlation as a proxy to study the quality of the reconstruction and to quantify the differences between the two methods. We consider the distance from every galaxy to its closest filament. Galaxies are typically not located at the exact centre of the filaments, and so this distance will have a dispersion around zero. In general, the main causes for this dispersion are the following:
        \begin{enumerate*}[nolistsep,label=(\alph*)]
                \item distribution of matter and galaxies along the whole size of the filament, not exclusively on the central spine,
                \item error in the location of the centre of the filament, and
                \item non-detection of actual filaments, so the galaxies in that filament will appear to be far from detected filaments.
        \end{enumerate*}
        
        The second point can be approximated with the estimated uncertainty of the detections. The third point is not significant in the full catalogue, but could be relevant if one takes a small subset, such as a strong threshold in the uncertainty of the detection. In our analysis, we consider the full catalogue.

\begin{figure}
                \centering
                \includegraphics[width=\columnwidth]{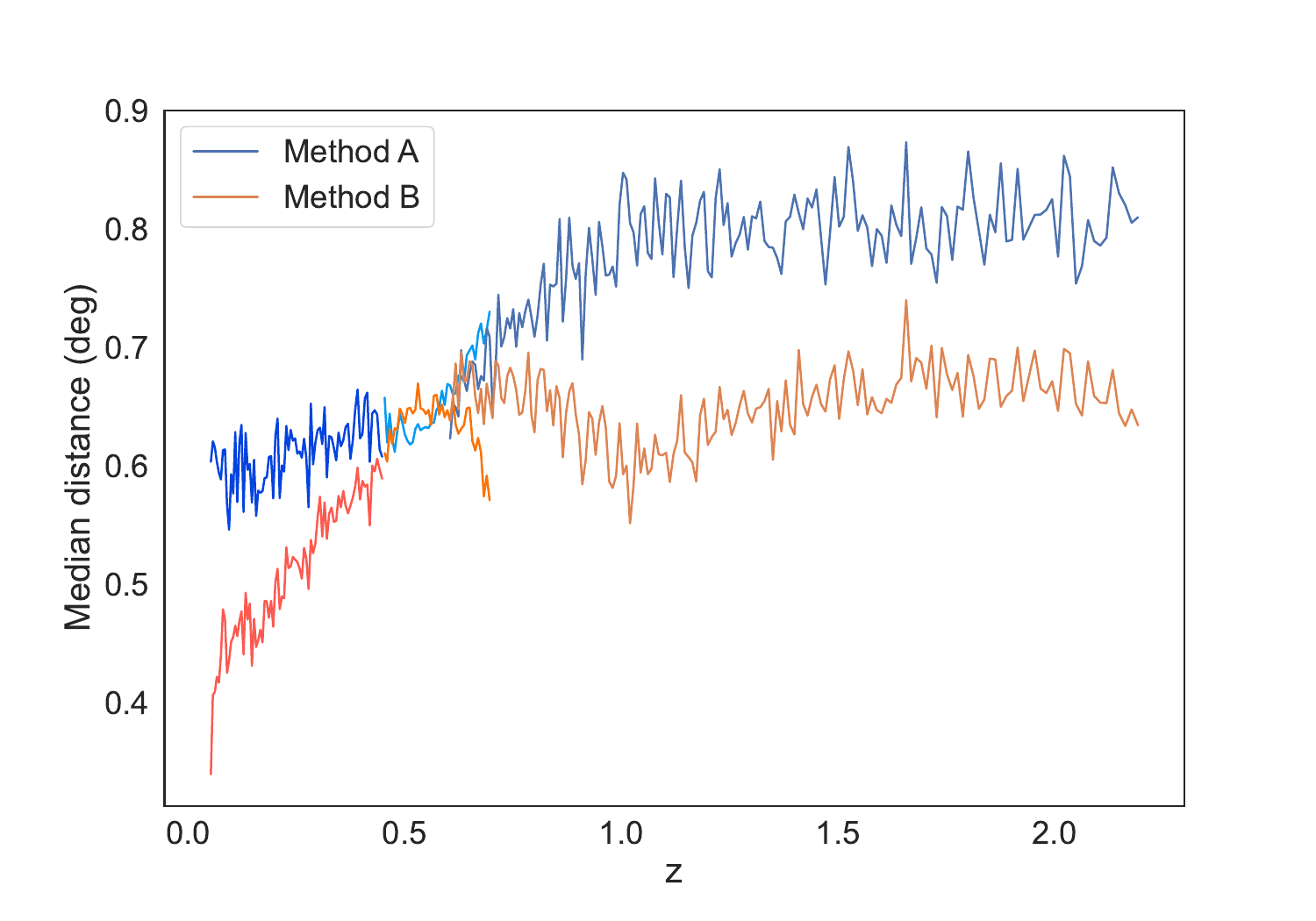}
                
                \includegraphics[width=\columnwidth]{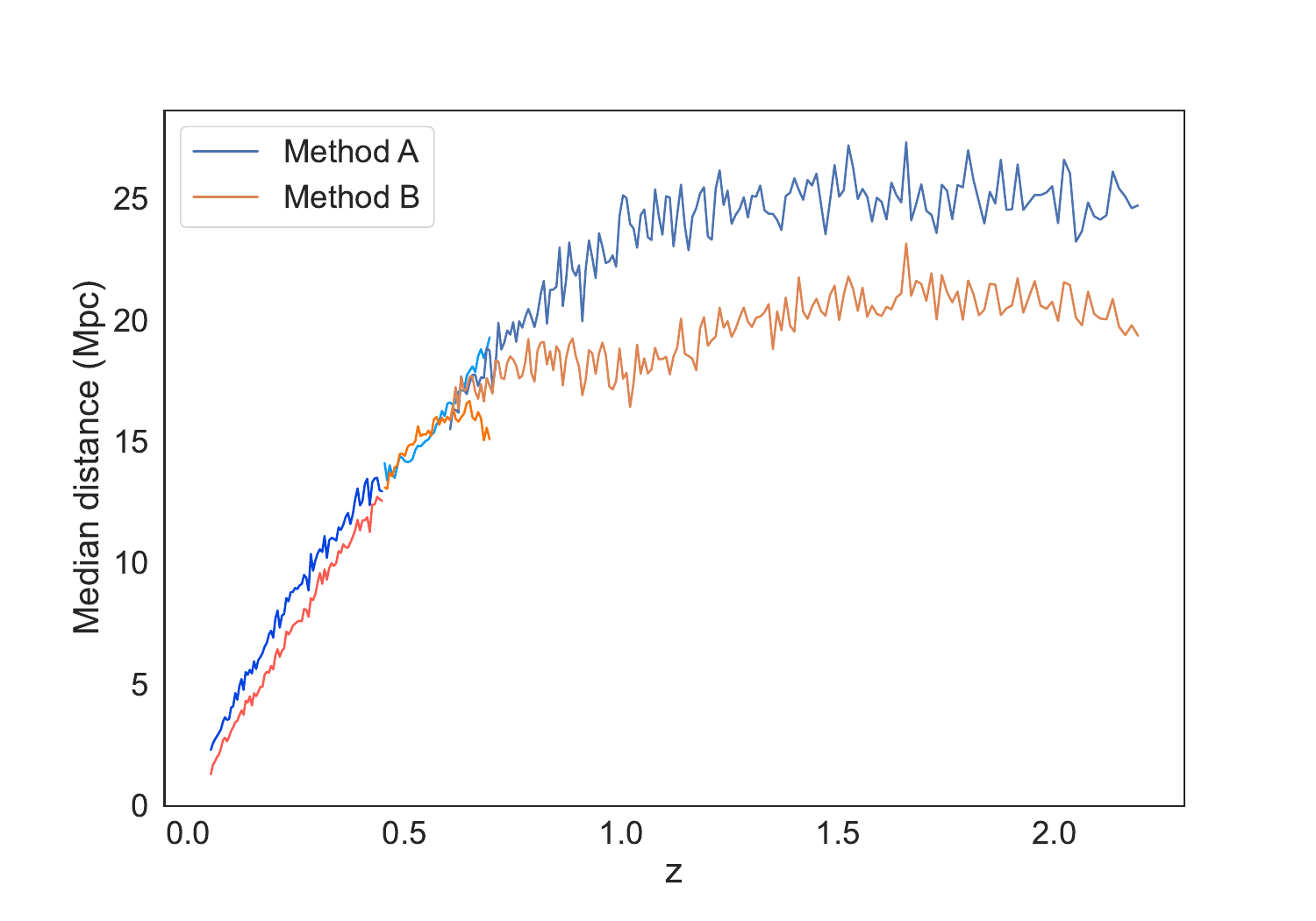}
                \caption{Mean distance between galaxies and their closest filament as a function of redshift. Blue corresponds to Method A, and orange corresponds to Method B. Different shades of colour correspond to the three different Blocks with different masks. Top: Angular distance in degrees; Bottom: Proper distance in megaparsecs.
                        \label{f:dists}}
        \end{figure}

        We compute the distance between galaxies and their closest filament for the two methods. A comparison of the median distance over redshift can be seen in \Cref{f:dists} both in degrees and proper distance. We choose to use the median as it is more robust in slices with a lower number of galaxies (especially at high and very low redshift). This distance is lower overall in Method B, partly due to the fact that it reports a higher number of filaments.
        
\begin{figure}
                \centering
                \includegraphics[width=\columnwidth]{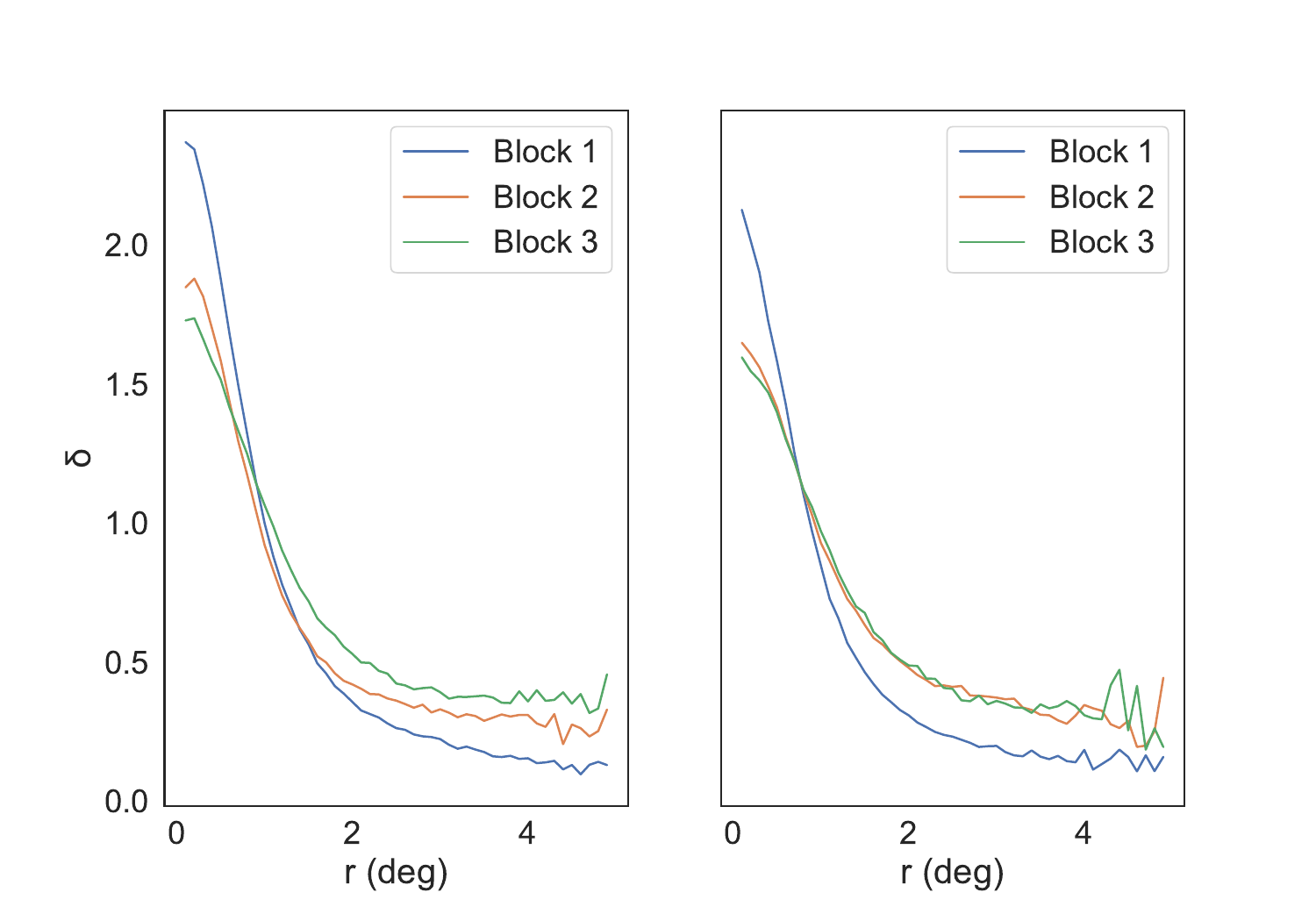}
                \caption{Average distribution of distances between galaxies and their closest filament divided by the same distribution for random points, as described in the text. Left: Method A. Right: Method B. Different colours correspond to the three different Blocks.
                        \label{f:ratio}}
        \end{figure}

        Another way to analyse the distance between galaxies and filaments is to look at their distribution compared to that obtained from a random distribution of points. In this case, we partially mitigate the effect induced by having a larger number of filaments in method B, as it also decreases the overall distance from random points to filaments. The number of galaxies at a distance $r$ from a filament, with respect to the number of random points at the same distance, $\delta(r)=\frac{n_{gal}(r)}{n_{ran}(r)}$, can be seen in \Cref{f:ratio} for both catalogues. Both catalogues present a similar ratio, although Catalogue A produces a slightly higher peak at a very close distance to the filaments.
        
        Both methods yield significant detection levels throughout the entire redshift range. This includes $z>1$, where data come from QSO detections only. The quality of the reconstruction in this range remains roughly constant. Distances from galaxies to filaments are significantly lower than for random points.

        This distance between galaxies and filaments can be understood as a typical width of the filament and is compatible with values in the literature. This quantity has been studied in hydrodynamical simulations by \citet{galarragaespinosa2020} using Illustris-TNG, Illustris, and Magneticum simulations. These authors observe that the radii of filaments are typically $\sim3$ to $\sim5$ Mpc at $z=0$. Although the radius definition is different and we do not reach $z=0$, we verified that our measurement in the lowest redshift bins is compatible with the measurement of these latter authors in simulations. This is also compatible with other simulations in the literature, such as \citet{colberg2005}, where the typical radius of a filament is between $1.5h^{-1}$ and $2h^{-1}$  Mpc.
        
        \subsubsection{Uncertainty of the detections}
        \label{ss:uncertainties}
        The second metric we use in the comparison is the estimated uncertainty of the detections. This is calculated using bootstrapping of the original galaxies, as explained in \Cref{alg:2}. This procedure estimates the variability in the location of the detected filaments, given in degrees, and is therefore a measure of the robustness of the results.
        
\begin{figure}
                \centering
                \includegraphics[width=\columnwidth]{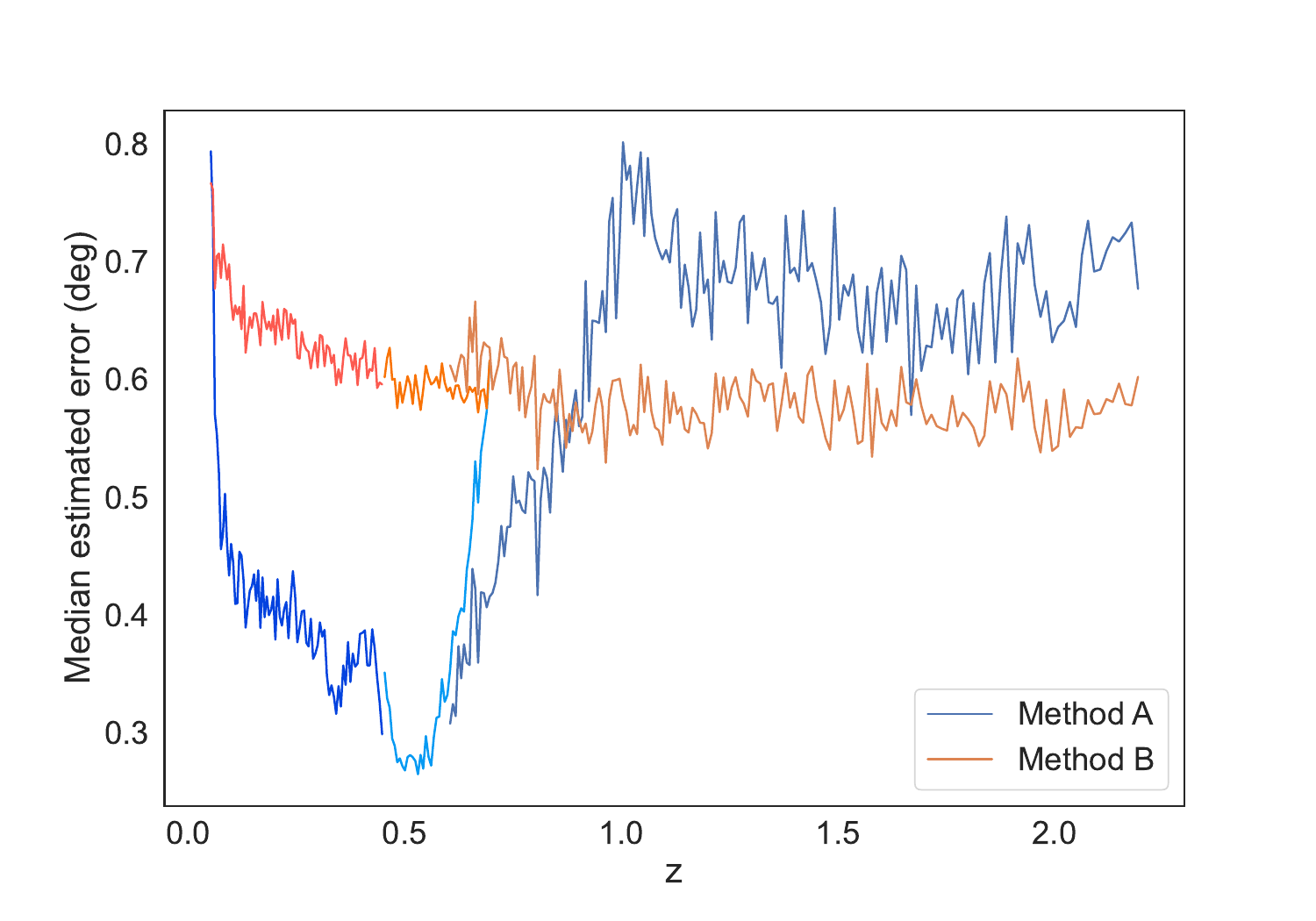}
                
                \includegraphics[width=\columnwidth]{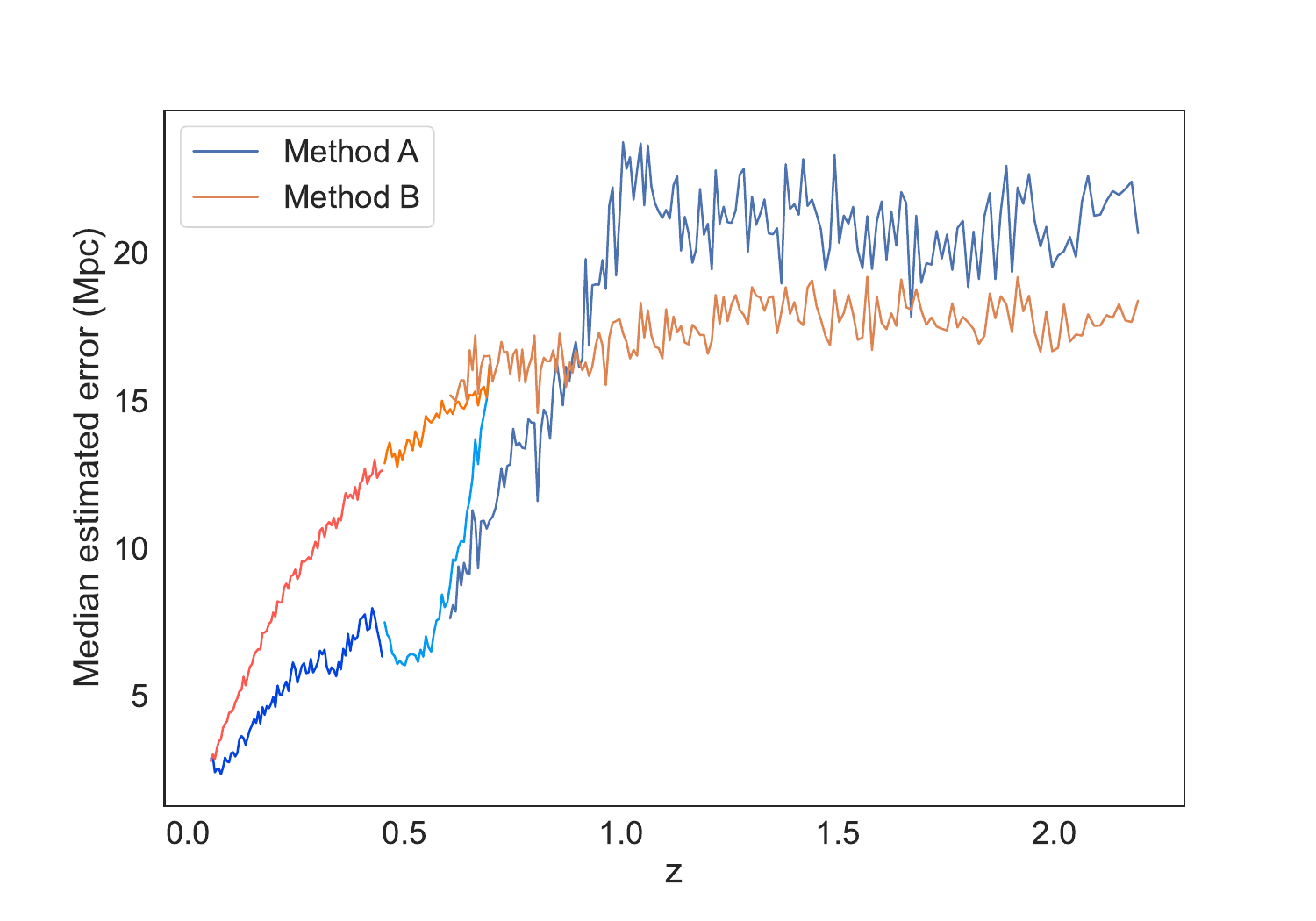}
                \caption{Mean uncertainty estimate for the detected filaments as a function of redshift. Blue corresponds to Method A, orange corresponds to Method B. Different shades of colour correspond to the three different Blocks with different masks. Top: Angular distance in degrees; Bottom: Proper distance in megaparsecs.
                        \label{f:errs}}
        \end{figure}

        We compute the median of the estimated uncertainty of detection. A comparison can be seen in \Cref{f:errs} in degrees and proper distance. The median uncertainty for Method B is approximately constant over $z$. As mentioned before, at low redshift, the higher median uncertainty with respect to method A is mainly driven by the additional population of detected filaments with higher uncertainty. This effect is also seen at higher redshift, but it is compensated by a much lower estimated uncertainty for the filaments found in both catalogues.

        It is important to note that, at low redshift, Method B yields a result of similar quality to Method A (as measured with galaxy--filament distances), even if Method A reports a lower estimated uncertainty. At higher redshift ($z>1$), Method B yields both lower distances and lower uncertainty estimates.

        \paragraph*{}
        We note that we expect most applications of these catalogues to require an additional threshold, for example on the limit of the estimated uncertainty required\footnote{This can be done manually on the catalogue or with the Python functions that we provide at \href{https://www.javiercarron.com/catalogue}{javiercarron.com/catalogue}}. After a cut of this kind (for example using Catalogue B while removing points with an estimated uncertainty over \SI{1}{\deg}), the described results would be slightly modified in the following way: (a) the median distance would increase overall by around \SI{0.05}{\deg}, (b) the peak in $\delta$ would increase to be very similar to the one in Catalogue A, and (c) the estimated uncertainty would be very uniform in $z$ at a value of $(0.55\pm0.02)$\si{\deg}. This is a consequence of the natural trade-off between the confidence of detections and the number of filaments detected that arises when imposing a threshold of this kind. Different applications will require different choices.

        \section{Comparison with previous work}
        \label{s:comparison}
        As stated above, this work follows a similar framework as the one described in \citet{chen2015, chen2016} (\YCC{}). In this section, we explore the improvements in the catalogues obtained by our implementation. 
        A first improvement concerns the data themselves. In this work we perform the reconstruction on a wider sky region encompassing both the north and south Galactic caps, while the \YCC{} catalogue is produced only in the north Galactic cap. We also include the eBOSS sample, which allows a deeper reconstruction in redshift. However, to make a fairer comparison of the methodologies, in the rest of the section we restrict to the redshift range given by \YCC{}: $0.05<z<0.7$, equivalent to Block 1 and Block 2 of our catalogues. These two Blocks are produced only with BOSS data. Additionally, in \YCC{} the authors use the NYU Value Added Catalogue \citep{blanton2005} to have more data in the lowest redshift bins. We decided against using these data in order to have a consistent dataset over the whole redshift range, produced only by SDSS. This means that comparisons at the lowest redshift bins ($z\lesssim0.2$) will be affected by the difference in the data being used, and so we do not consider them in the comparison. 
        
        We use our Catalogue A in the comparisons, as it is the most similar to the methodology followed to produce \YCC{}. The main difference is the spherical treatment of the sky; other minor differences are the binning strategy ($20$ Mpc instead of $\Delta z = 0.005$) and slightly different criterion for the smoothing kernel. In \Cref{ss:comp} we compare the results obtained using our two methods: Method A, our implementation of the SCMS algorithm, and Method B, a version boosted with a machine learning approach.

        \subsection{Uncertainty of the detections}
        The first metric we use to compare our Catalogue with the \YCC{} catalogue is the estimated uncertainty for each point in a filament; this is the same metric we used to compare our two catalogues in \Cref{ss:uncertainties}. This is calculated using bootstrapping of the original galaxies, as explained in \Cref{alg:2}. This estimates the variability in the location of the detected filaments, given in degrees and is therefore a measure of the robustness of the results.
        
\begin{figure}
                \centering
                \includegraphics[width=\columnwidth]{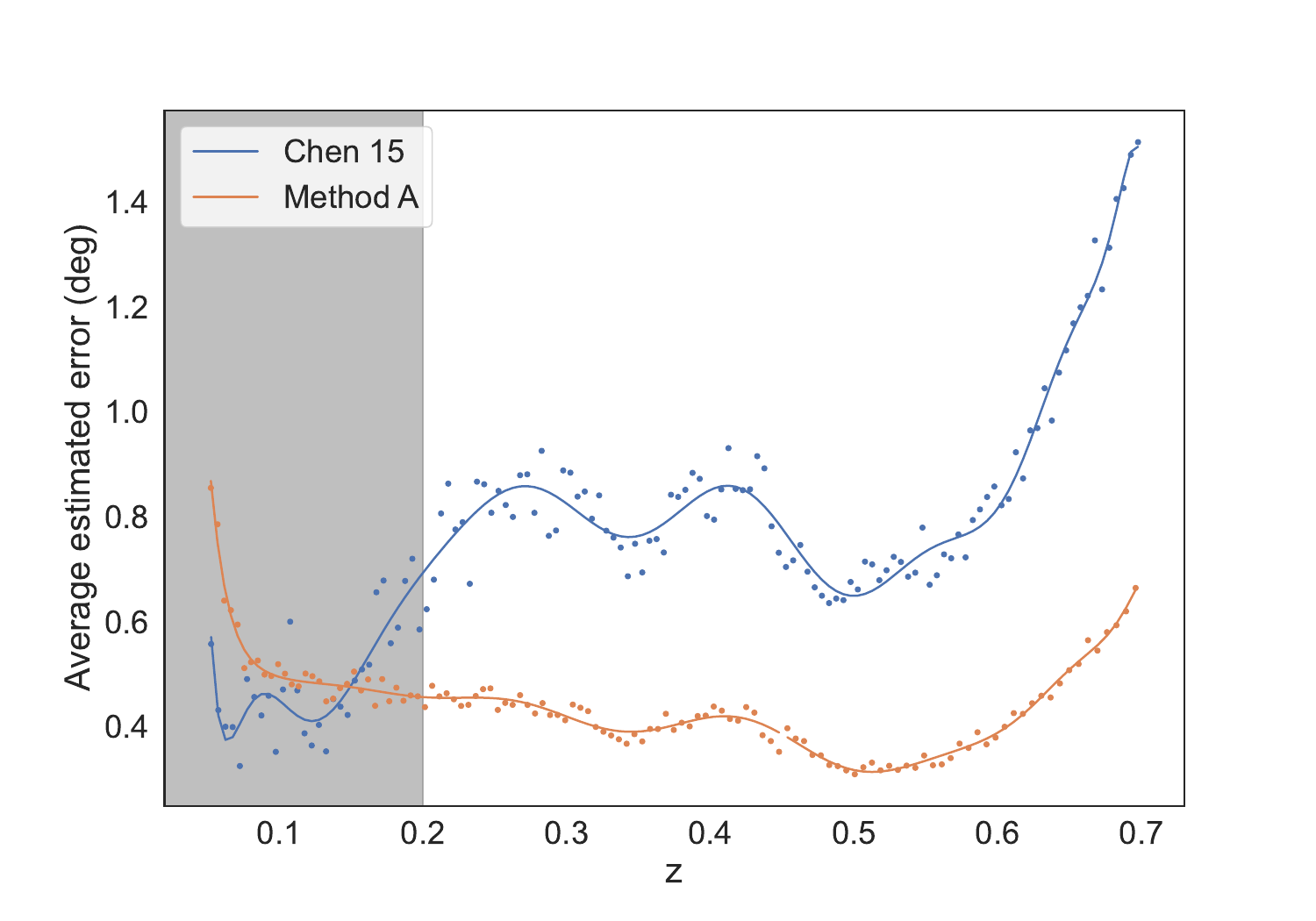}
                \caption{Average estimated uncertainty of the detected filaments as a function of redshift. Orange corresponds to our Method A, blue corresponds to \YCC{}. For $z<0.2$ (grey band), filaments are reconstructed from different data in the two catalogues. Lines are non-linear fits shown for visualisation only.
                        \label{f:yccerr}}
        \end{figure}

        For each redshift, we compute the average uncertainty estimate. This can be seen in \Cref{f:yccerr}. We observe that the estimated uncertainty is lower in our catalogue over the entire redshift range,  by a factor of approximately two. It is also more uniform with redshift. We stress that the uncertainty was estimated using the same method adopted in \YCC{} (\Cref{alg:2}), and therefore the difference we find comes from the reconstruction of filaments.

        \subsection{Distance from galaxies to filaments}
        The second metric we use to compare the catalogues is the distance between galaxies and their closest filament. As explained in \Cref{ss:distance}, this quantity is mainly determined by the physical width of the filaments and the error in their detected positions.
        
        \begin{figure}
                \centering
                \includegraphics[width=\columnwidth]{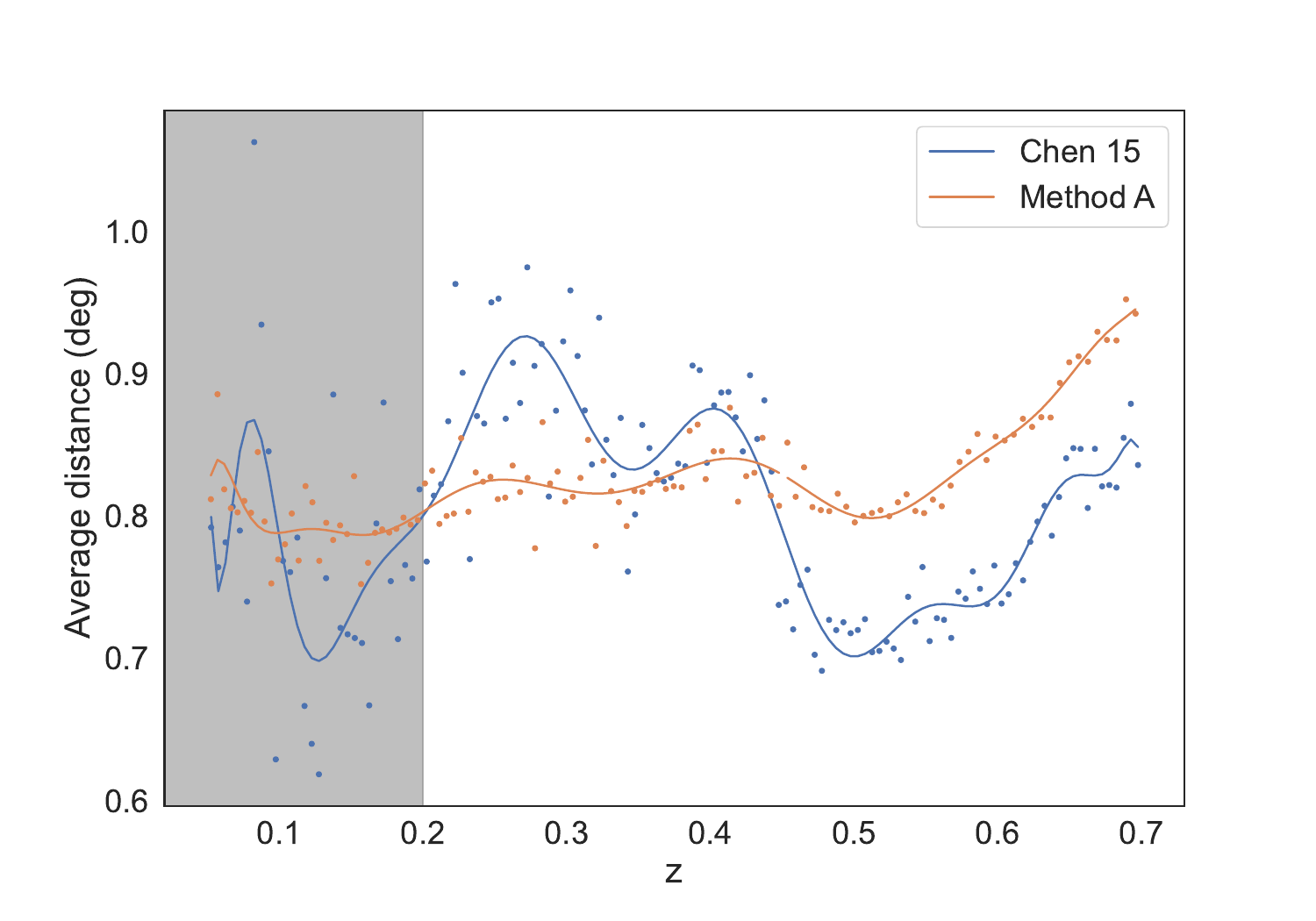}
                \caption{Average distance between galaxies and their closest filament as a function of redshift. Orange corresponds to our Method A, blue corresponds to \YCC{}. For $z<0.2$ (grey band) filaments are reconstructed from different data in the two catalogues. Lines are non-linear fits shown for visualisation only.
                        \label{f:yccdist}}
        \end{figure}
        
        We compute the distribution of these distances and obtain the average for each redshift slice. The comparison of average distances can be seen in \Cref{f:yccdist}. Our catalogue produces lower distances between galaxies and filaments for $z<0.45$. \YCC{} presents a sharp decrease at redshift $z\approx0.45$, corresponding to the transition from LOWZ data to CMASS data. As we explore in \Cref{s:iso}, this is mainly driven by artefacts at high latitudes, where the flat sky approach is less accurate. We observe a smoother curve with less dispersion in our catalogue, corresponding to more consistent detections across redshifts.

        Additionally, we can compare the average distances between filaments and galaxies with the estimated uncertainties for the centre of the filaments. These two quantities are comparable for \YCC{}, whereas in our Catalogue A the uncertainty is lower by a factor of  approximately two. This means that, in our case, the dispersion in distances is not dominated by the uncertainty in the location of filament centres, although this effect cannot be neglected.

        \subsection{Isotropy}
        
        \label{s:iso}
        
        One of the main differences between our work and \YCC{} is the full treatment of the spherical geometry of the sky. The catalogue in \YCC{} is obtained by approximating the observed area to a flat sky using spherical coordinates as euclidean. As a consequence, there is some risk that results could show anisotropic features at high latitudes. On the other hand, as we do not apply such an approximation, the algorithm is expected to behave isotropically.         
        
        We test this expectation by considering two stripes at different latitudes\footnote{We are always working in an equatorial reference system, so the latitude of the sphere corresponds to the declination $\delta$.}:
        \begin{equation}
                \begin{split}
                        45<& \: lat < 65 \, \deg,\\
                        0<& \: lat < 20 \, \deg.
                \end{split}
        \end{equation}
        
\begin{figure}
                \centering
                        \centering
                        \includegraphics[width=\columnwidth]{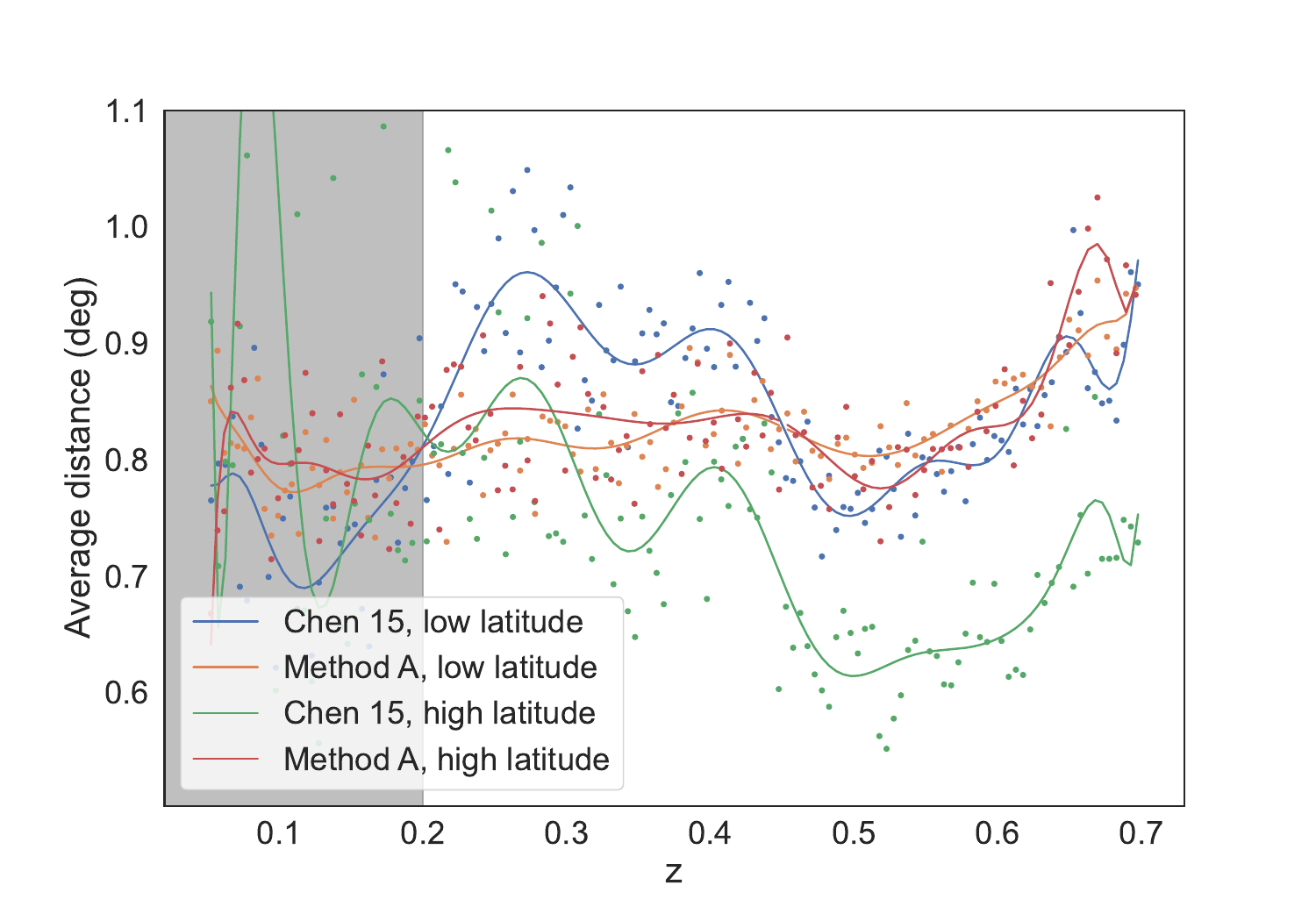}
                        \centering
                        \includegraphics[width=\columnwidth]{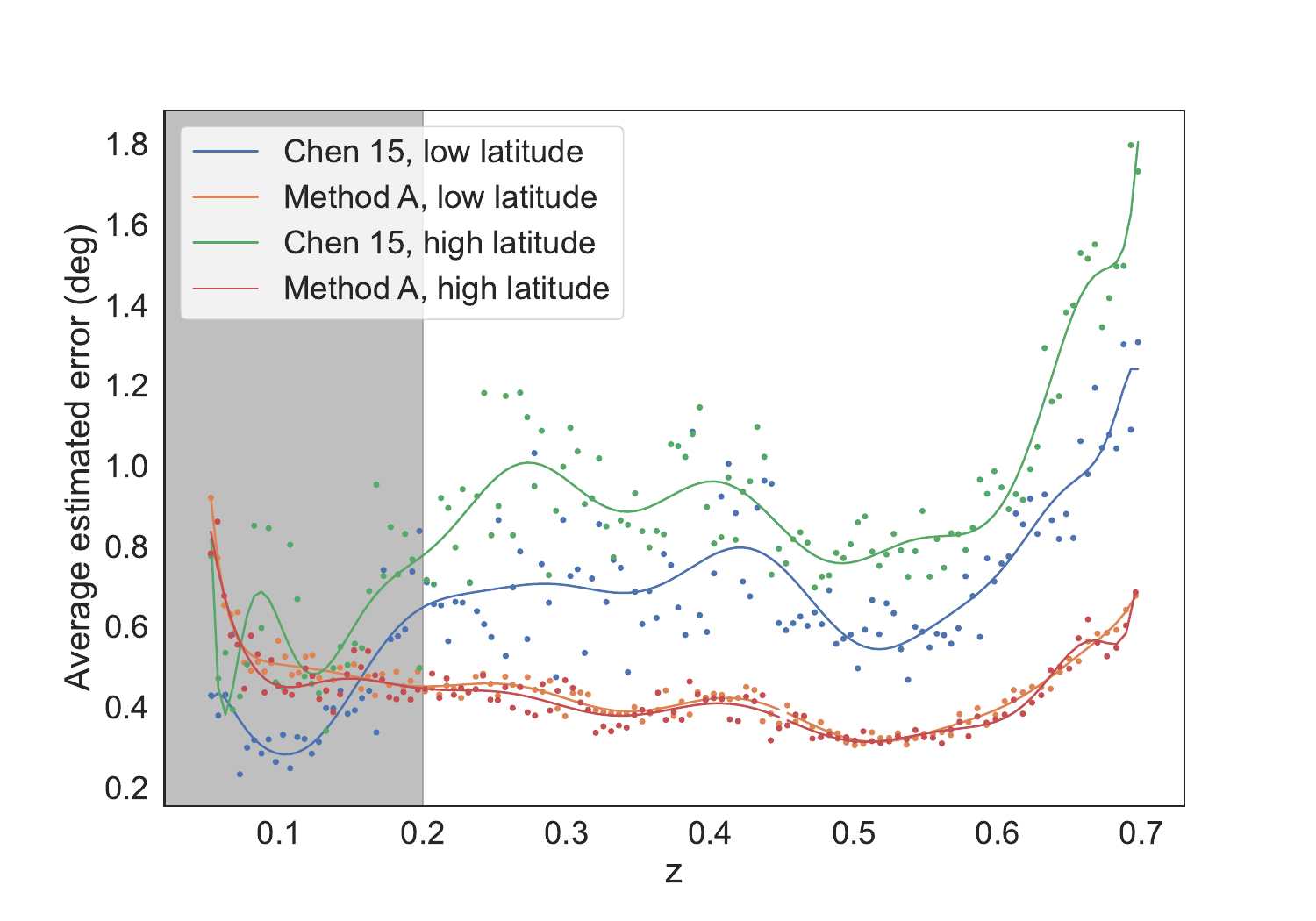}
                \caption{Top: Average distance from galaxies to their closest filaments, in degrees. Bottom: Average estimated uncertainty on the location of the centre of the filaments, in degrees. Both metrics are calculated separately for high- and low-latitude filaments. For $z<0.2$ (grey band), filaments are reconstructed from different data in the two catalogues. Lines are non-linear fits shown for visualisation only.
                        \label{f:isotropy}}
        \end{figure}

        We look at the same metrics discussed above, restricting the analysis to these stripes. The estimated uncertainty and the galaxy--filament distance can be seen in \Cref{f:isotropy}. The first thing to notice is that our catalogue reports highly compatible results at high and low latitudes, both for the estimated uncertainty and the galaxy--filament distance. Our algorithm is performing isotropically, as expected.
        
\begin{figure*}
                \centering
                \begin{subfigure}{0.45\textwidth}
                        \centering
                        \includegraphics[width=\textwidth]{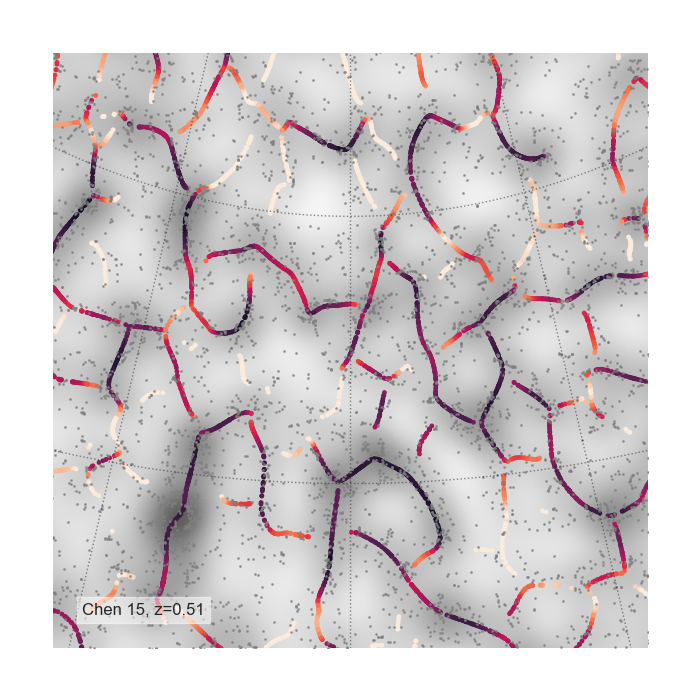}
                \end{subfigure}
                \begin{subfigure}{0.08\textwidth}
                        \centering
                        \includegraphics[width=\textwidth]{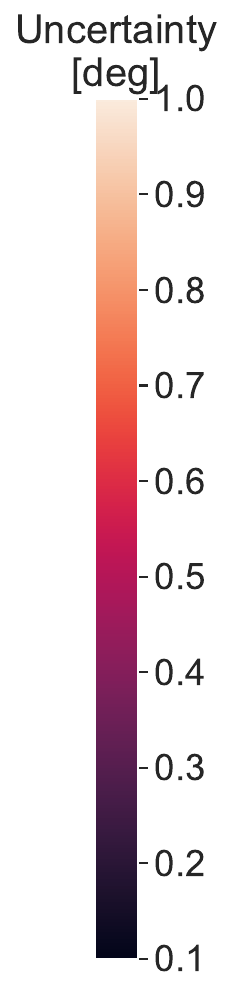}
                        \vspace{0.5cm}
                \end{subfigure}
                \begin{subfigure}{0.45\textwidth}
                        \centering
                        \includegraphics[width=\textwidth]{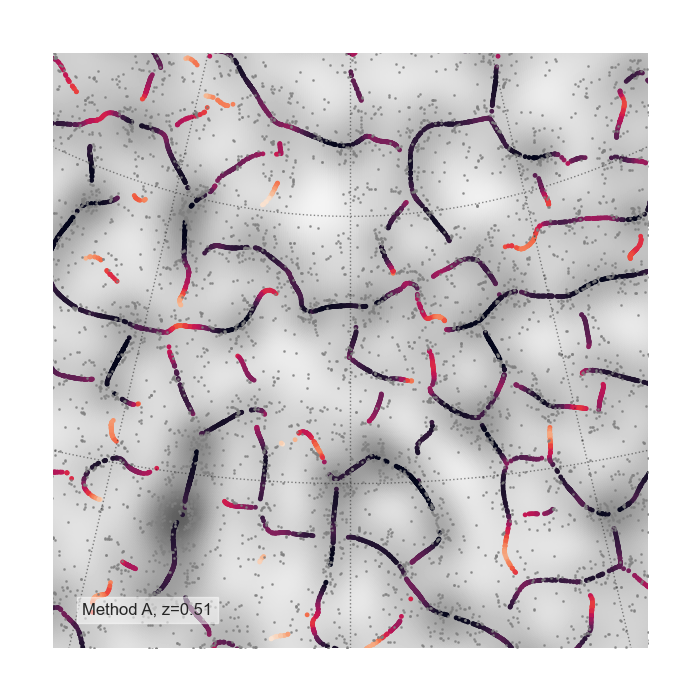}
                \end{subfigure}
                
                \caption{Example of obtained filaments at $z=0.510$ for \YCC{} (\textit{left}) and our result (\textit{right}).
                        The colour of the points represents the uncertainty estimate, in degrees. Grey points correspond to galaxies, while the background map represents the galaxy density smoothed at \ang{3} for visualisation purposes. The visualisation is a gnomonic projection around the point ($RA=140$, $\delta=40$) with a side size of \ang{60}. Meridians and parallels are both represented in steps of \SI{20}{\deg}.
                        \label{f:iso_highz}}
        \end{figure*}

        The \YCC{} catalogue on the other hand shows a slightly different behaviour for results at higher and lower latitudes. An example of this difference can be seen in \Cref{f:iso_highz}, where we compare the filaments obtained at $z=0.510$ for both catalogues. It can be seen that the filaments at low latitudes are similar, while there is a larger discrepancy at high latitudes. In that region, \YCC{} produces a larger number of filaments, with a preferred vertical orientation. This is an artefact due to the flat sky approximation and indeed is hidden when looking at the filaments in a flat sky, because this region gets dilated (e.g. see Figure 4 in \citealp{chen2016}). This anisotropic feature artificially reduces the distances from galaxies to filaments in this region compared to low-latitude results with no anisotropic effects. This reduction can be up to $25\%$  in the average distance. Heuristically, it seems that the intrinsic physical meaning of the derivative in the $\phi$ direction is altered by flat sky projection effects at high latitudes and this has an effect on the features of the algorithm.

        Similarly, the estimated uncertainty reported by \YCC{} is also affected by this anisotropy. Estimated uncertainties are calculated as an average of distances between real filaments and simulated filaments (as explained in \Cref{alg:2}). However, distances calculated in flat space are artificially increased at high latitudes. This implies an increased value of the estimated uncertainty for this region. Indeed, this is the observed effect, as presented in \Cref{f:isotropy}.
        
        This effect can be quantified by defining $\Delta dist$ and $\Delta unc$ as the average difference between values at high and low latitudes of the distance and the estimated uncertainty. For Method A we have the following results:
        \begin{equation}
                \begin{split}
                        \Delta dist &= 0.005\,\deg\\
                        \Delta unc &= -0.01\,\deg,
                \end{split}
        \end{equation}
        
        while for \YCC{} we obtain the following:
        \begin{equation}
                \begin{split}
                        \Delta dist &= -0.075\,\deg,\\
                        \Delta unc &= 0.23\,\deg.
                \end{split}
        \end{equation}
        
        We note that this anisotropy is only relevant at high latitudes. At low and intermediate latitudes, our catalogue reports similar results to those of \YCC. Therefore, this effect does not invalidate the multiple results in the literature based on this catalogue. If anything, we expect these results to be more significant when this anisotropy is corrected.

        \section{Validation of the catalogue}
        \label{s:validation}
        In this section, we evaluate the quality of the reconstruction in several respects: we check that the results are isotropic, that the catalogue is strongly correlated to independent galaxy cluster catalogues, that the results are stable even if new data are included, and that the length distribution of filaments is compatible with other works in the literature.
        
        \subsection{Isotropy}
        An interesting aspect related to the isotropy of our catalogue is the result at both hemispheres. For $0.05<z<0.7$ (i.e. Block 1 and 2), we have data at both the southern and northern hemispheres. The sky fractions\footnote{The sky fraction is defined as the observed area over the total sky area; the maximum value of $f$ for a hemisphere is therefore $0.5$.} are $f_S=0.054$, $f_N=0.132$ for Block 1, and $f_S=0.054$, $f_N=0.165$ for Block 2. As a sanity check, we test that the results at both hemispheres are similar.
        
        On each redshift slice and each hemisphere, we compute the mean distance from galaxies to filaments, and the mean estimated uncertainty. We then compute the difference between the metrics at both hemispheres (north minus south). The average differences for Method A are
        \begin{equation}
                \begin{split}
                        \Delta dist &= 4.4 \cdot 10^{-5}\,\deg,\\
                        \Delta unc &= 7.0 \cdot 10^{-3}\,\deg,
                \end{split}
        \end{equation}
        
        and the average differences for Method B are
        \begin{equation}
                \begin{split}
                        \Delta dist &= 5.7 \cdot 10^{-3}\,\deg,\\
                        \Delta unc &= 5.4 \cdot 10^{-2}\,\deg.
                \end{split}
        \end{equation}
        
        We see that none of these differences are significant: all of them are below $1$ arcminute, except the difference in estimated error with Method B, which is \ang{;3.2;}. Additionally, we do not observe trends with redshift.
The algorithms is performing isotropically, yielding compatible results at both hemispheres.

        \subsection{Galaxy clusters}
        
        According to the evolution of the cosmic web, the largest overdensities are located in the cosmic halos, which are usually located at the intersection of several filaments. These overdensitites can seed the formation of galaxy clusters. Therefore, known galaxy clusters are expected to be located at a small distance from cosmic filaments \citep{hahn2007,aragoncalvo2010}.
        
        We test this expectation with two catalogues of galaxy clusters available in the literature, listed below. Both catalogues were built using the Sunyaev-Zel'dovich effect as a tracer. Therefore, the detection of these galaxy clusters is completely independent of the SDSS data we use to detect the filaments. The catalogues are the following:
        \begin{itemize}
                \item Planck galaxy cluster catalogue: \citet{planckcollaboration2016a}. This catalogue presents a total of $1653$ clusters, mainly at $z<0.6$, corresponding to our Blocks 1 and 2. They are reported to have a $S/N\geq4.5$. They are distributed over the full sky; a total of $301$ clusters are located within the region where we extract the filaments ($254$ for Block 1 and $47$ in Block 2).
                
                \item Atacama Cosmology Telescope (ACT) galaxy cluster catalogue: \citet{hilton2020}. This catalogue presents a total of $4195$ clusters, with $S/N\geq4.0$. This catalogue covers a restricted sky area, as it presents clusters with latitudes up to \ang{20}. Therefore, there is some overlap with the footprints for our Blocks 1 and 2. Unfortunately, there is no overlap with Block 3. In order to minimise possible systematic errors, we select the subsample of clusters with spectroscopic determination of redshift. The number of selected clusters in the footprint of the filaments is $951$ ($479$ in Block 1 and $472$ in Block 2).
        \end{itemize}
        
\begin{figure*}
                \centering
                \includegraphics[width=0.8\textwidth]{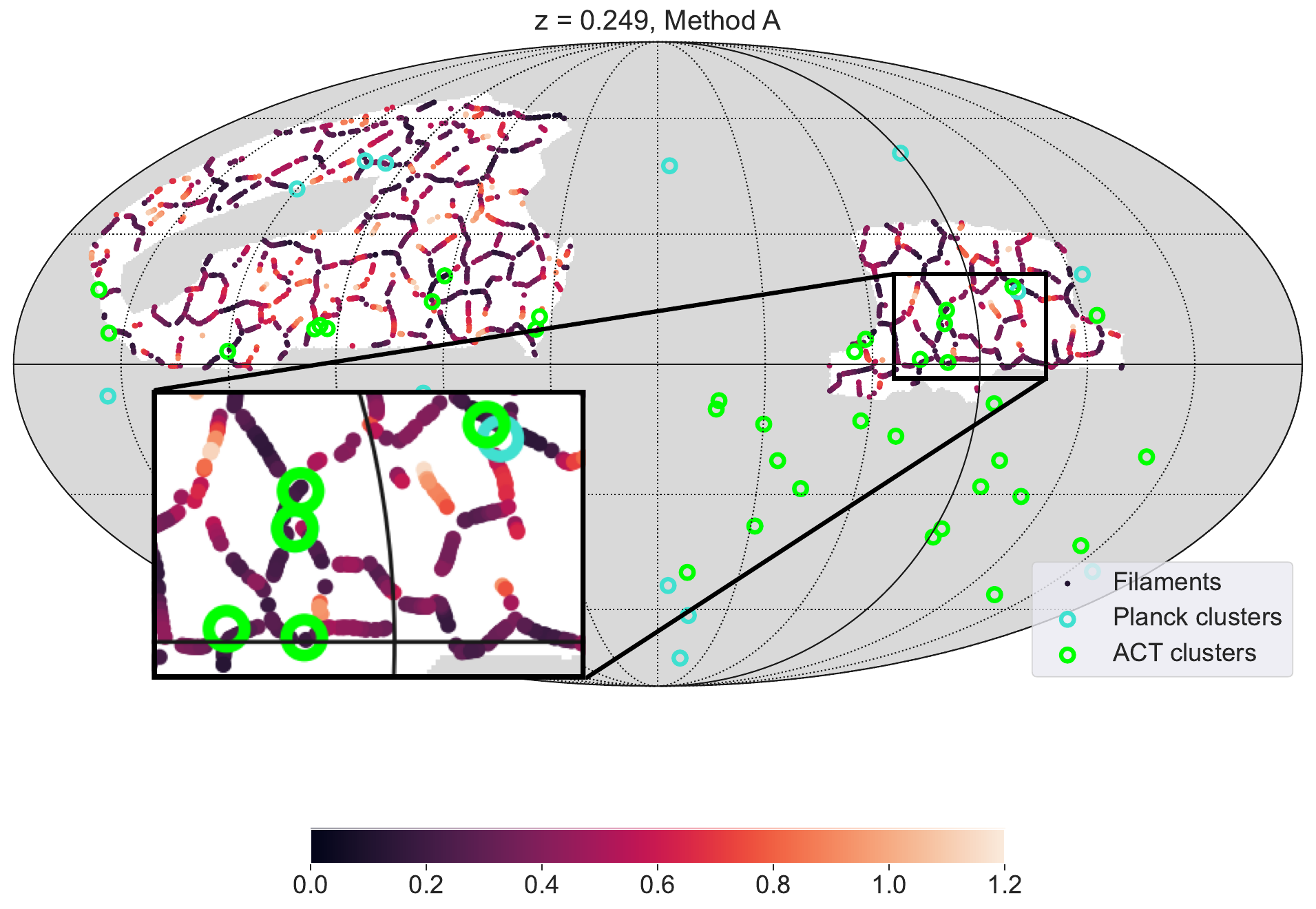}
                \vspace{1cm}
                
                \includegraphics[width=0.8\textwidth]{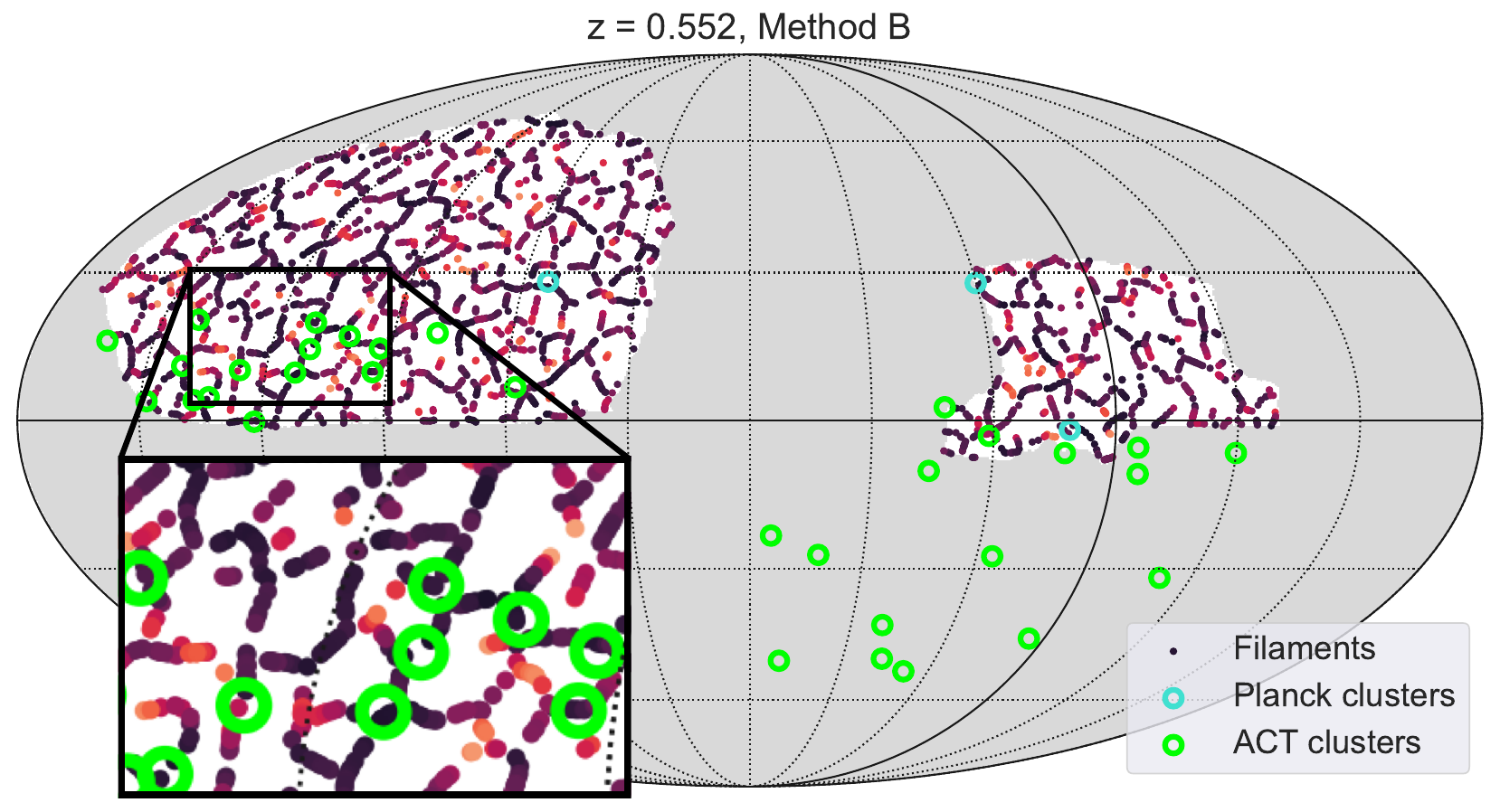}
                \caption{Example of filament maps at different redshift slices with the two methods. The colours of the filaments represent the estimated uncertainty of the detection, in degrees. Independently-measured galaxy clusters from Planck and ACT are also represented. The grey area is excluded from the catalogue due to lack of data and border effects.
                        \label{f:clusmap}}
        \end{figure*}

        Two examples of the location of clusters and filaments can be found in \Cref{f:clusmap} for $z=0.249$ (Method A) and $z=0.552$ (Method B). All clusters are found to be close to filaments and in particular they tend to be close to intersections of filaments. We observe this trend for both catalogues and throughout the entire redshift range. In some rare instances, a cluster seems to be far from any filament; this can be seen in the bottom image of \Cref{f:clusmap} for the cluster at $lat=$\ang{17.5}, $lon=$\ang{169.0}, apparently located in a cosmic void. A closer examination of this individual cluster reveals that its redshift ($z=0.553$) is very close to the lower limit of the slice ($z=0.552$). Indeed, it is located in the intersection of filaments in the previous redshift slice, hinting at a possible small error in its redshift determination.
        
\begin{figure}
                \centering
                \includegraphics[width=\columnwidth]{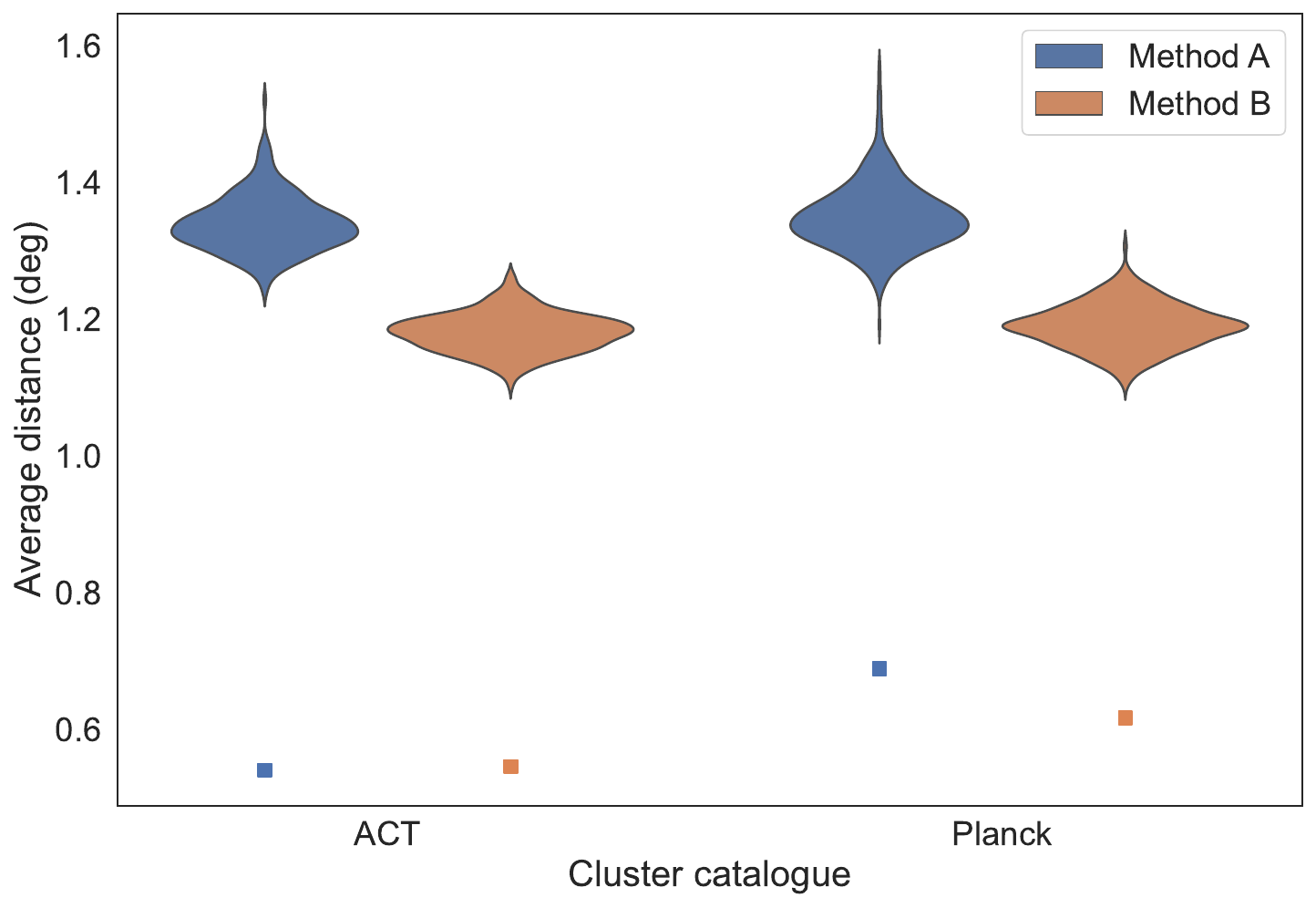}
                \caption{Distribution of the average distance between clusters and filaments for random realisations and real data. In the y-axis, the average distance, in degrees; different histograms for the two cluster catalogues and methods are represented. The average distance for real data is represented with squares.
                        \label{f:clus}}
        \end{figure}

        In order to test the statistical significance of the correlation between these clusters and our filaments, we produce $500$ random mocks for each cluster catalogue. For each simulation, we take all the clusters located in the footprint of our filaments and randomly place them in a different location, keeping their original redshift. The location is sampled uniformly from the intersection between the cluster catalogue footprint and the filaments footprint. We compute the average distance from these new points to the filaments. We repeat this calculation for the actual data and see how different it is from the random realisations. We can see the result in \Cref{f:clus}. It can be seen that the actual clusters are much closer to the filaments than the random catalogue. In particular, the average distance with actual data is between $14\sigma$ and $23\sigma$ lower than random realisations.
        
        The average distance between clusters and the closest filaments is \ang{0.54} for ACT and \ang{0.62} for Planck. There is therefore a strong correlation between the filament catalogues presented in this work and cluster catalogues in the literature. We emphasise that the detection of these clusters is obtained with a different physical observable and they are independent from our data.

        \subsection{Stability when faced with  new data} 
        
        One question that arises when analysing the results is how robust these filaments are with respect to new data. If a future survey is able to detect more galaxies at a certain redshift, we would like to know whether we would obtain a similar set of filaments. Equivalently, this can be seen as the predictive power of the filaments: whether new observations of galaxies will be located in the regions predicted by the detected filaments.
        
        We can test this exact situation in the redshift range of $0.6<z<0.7$. In this range, we have data coming from BOSS that we use in our Block 2, but we also have more recent data coming from eBOSS; we use an aggregation of both in the Block 3. The two Blocks overlap in this redshift range, and so we can test how much the filaments differ when adding the more recent eBOSS data. We note that eBOSS sky coverage is smaller, so we limit this comparison to the common sky region.
        
        In order to compute the difference, we take the filaments obtained with BOSS data only and consider an area around them with a radius of \ang{1}. We then take the filaments obtained with BOSS+eBOSS data and compute the fraction of filament points that lay inside this region. This procedure is done for all redshift slices in $0.6<z<0.7$. Slices at lower redshift have a higher percentage of common galaxies; that is, new data coming from eBOSS are less important here, and we expect to find similar filaments. At higher redshift, the majority of galaxies come from new data, and we expect the filaments to differ more. 
        
\begin{figure}
                \centering
                \includegraphics[width=\columnwidth]{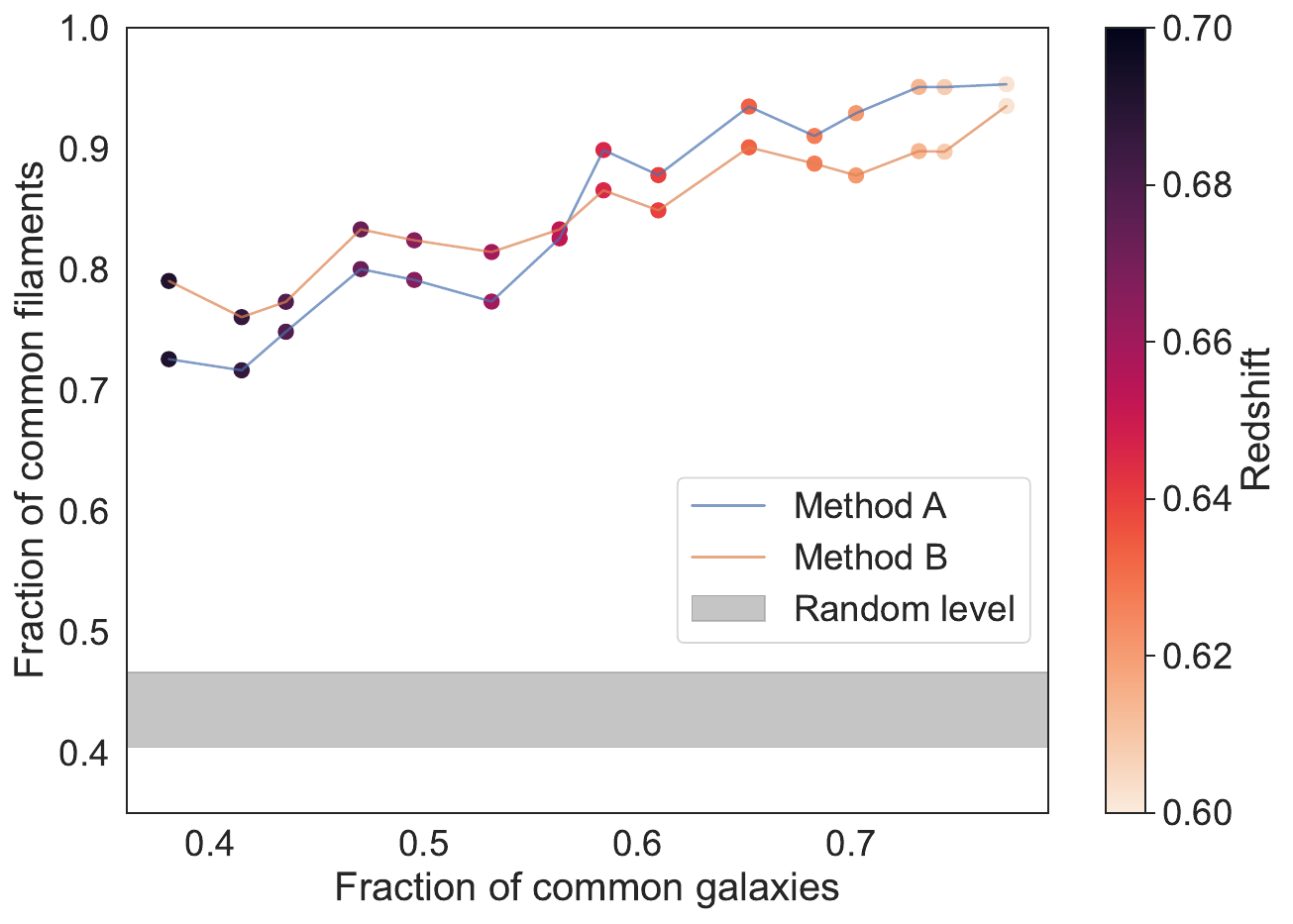}
                \caption{Stability of the results to new data. On the y-axis, the fraction of filaments found using all data (older and newer) which were also found using only older data. On the x-axis, the fraction of galaxies corresponding to older data. Each point corresponds to the result for a single redshift slice. See text.
                        \label{f:stability}}
        \end{figure}

        We can see the fraction of common filaments as a function of the fraction of common galaxies in \Cref{f:stability}. Each point corresponds to a single redshift slice, and the two curves correspond to Method A and B. The grey region corresponds to the expected results when carrying out this procedure with uncorrelated slices. The fraction of common filaments (alternatively, the fraction of filaments that are correctly predicted by older data) is between $72\%$ and $96\%$, even when the fraction of shared galaxies goes below $40\%$.
        
        This means that new observations in this range are unlikely to change the results significantly: duplicating the number of galaxies only produced $25\%$ more filaments. New data may have a greater effect if a different tracer is used, or with further improvements to the algorithms.

        \subsection{Length distribution of the cosmic filaments}

        The study of the topology of the cosmic web is highly non-trivial and is an active topic in the literature; see, for example \citet{neyrinck2018} and \citet{wilding2020}. This is in part due to the difficulty of defining {a single filament}. In this work, we  use a definition of filaments that is based on the ridge formalism, which is very useful to find the network of filaments, but does not provide information about individual filaments (see \Cref{ss:definition}).
        
        For this section, in order to provide example, we consider that every intersection of the web is the starting point of a filament, regardless of the angle or density of such intersections. The filament ends at another intersection or an end point. We note that this kind of definition is sensitive to the depth of the detection: detection of fainter and shorter filaments will of course increase the number of shorter filaments, but it will also increase the number intersections and split longer filaments into shorter ones.
        
        For a first exploration of the matter, we find the individual filaments on each redshift slice as follows: \begin{enumerate*}[label=\textit{(\alph*})]
                \item a binary HEALPix map of $nside=256$ is created with value of 0 everywhere,
                \item the values of the pixels close\footnote{At a distance of $4$ pixels or fewer from the filament point.} to a filament point in the catalogue are changed to $1$,
                \item the skeleton of this map is extracted\footnote{This is done with the software \texttt{Scikit-image} \citep{walt2014}, which uses the algorithm described in \citet{zhang1984}, iteratively removing pixels from the borders without creating gaps or disconnections.},
                \item pixels in the skeleton are divided into filaments, ends, or intersections depending on the number of neighbours, and
                \item filaments are considered to be the connected set of pixels between intersections or ends.
        \end{enumerate*}
        
        \begin{figure}
                \centering
                \includegraphics[width=\columnwidth]{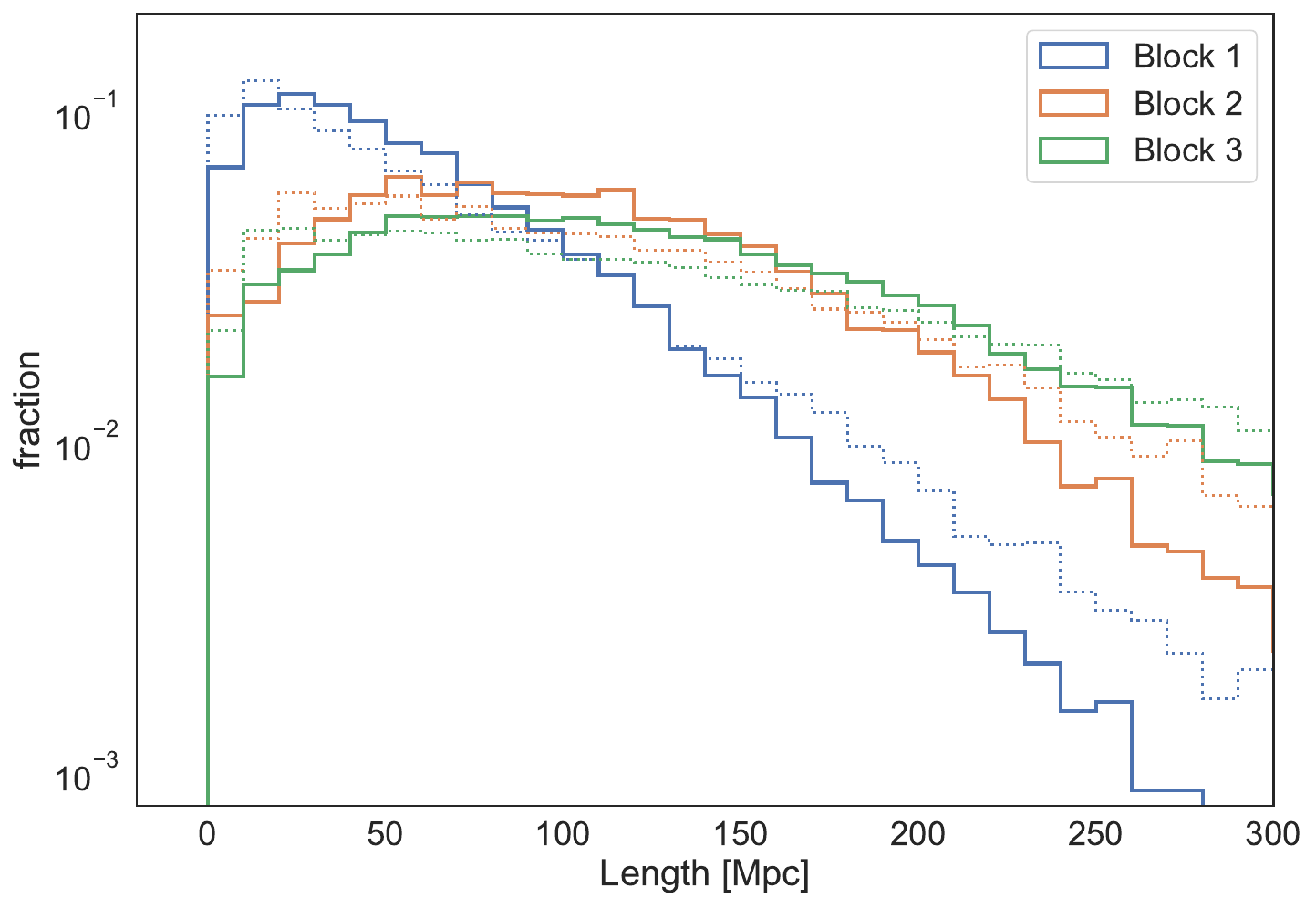}
                \caption{Distribution of the lengths of the filaments in the catalogues, for the three different Blocks (Block 1: $0.05<z<0.45$, Block 2: $0.45<z<0.7$, Block 3: $0.6<z<2.2$). The dotted line represents the results when we impose an additional threshold on the catalogue, considering only points with uncertainty of the detection under \ang{0.8}.
                        \label{f:length}}
        \end{figure}
        
        Using this procedure, we obtain the distribution of filament lengths of filaments in Catalogue B, which can be seen in \Cref{f:length}. It is characterised by a redshift-dependent peak followed by a power-law decrease. This distribution is also observed in \citet{malavasi2020} (see their Figs. 18-22); the length distribution found by these latter authors is almost exactly the same as our result for Block 1, even though the method they use to find filaments is completely different (i.e. $3D$ reconstruction with the DisPerSE scheme). For Blocks 2 and 3, at higher redshifts, we notice that there are fewer shorter filaments; this could be an observational effect (shorter filaments are more difficult to detect at higher redshifts). We note that a similar effect has been observed in simulations: less densely populated simulations report longer filaments \citep{galarragaespinosa2020}; although these simulations report shorter filaments overall. Further investigation is needed to further investigate these trends.
        
        Finally, we also report the length distribution when imposing an additional threshold to the catalogue. The fainter lines in \Cref{f:length} represent the result when considering only points with associated uncertainty lower than \ang{0.8}. As expected, the number of shorter filaments decreases, which decreases the number of intersections, increasing the number of longer filaments. We note that this threshold provides a similar result to that produced using Method A because this method is less sensitive to short filaments, as explained in \Cref{ss:comp}.

        \section{Conclusions}
        \label{s:Conclusions}
        
        In this work we present a new catalogue of cosmic filaments found with the SDSS Data Release 16. We implemented a version of the Subspace Constrained Mean Shift algorithm on spherical coordinates and applied it to thin redshift slices in order to obtain a tomographic reconstruction of the cosmic web. We boosted the algorithm with machine learning techniques, which improves the results at higher redshifts. Our main results can be summarised as follows:
        
        \begin{itemize}
                \item We report a new public catalogue of cosmic filaments, which inherits the sky footprint of SDSS, and its redshift coverage from $z=0.05$ up to $z=2.2$. All detected points of the cosmic web come with an associated uncertainty.
                \item We study the distances from galaxies to cosmic filaments, and the uncertainty associated with the filament detections, both as a function of redshift. We use these as metrics to assess the goodness of the reconstruction. Both metrics remain within reasonable bounds even at high redshift.
                \item We build upon the SCMS algorithm, which has already proven useful for filament detection and cosmological applications \citep{chen2015}. We show improvements in the isotropy of our results and we extend the coverage to higher redshift and the southern Galactic hemisphere thanks to more recent data.
                \item We additionally boost the algorithm with a machine learning algorithm that `learns' the characteristics of filaments from the most populated redshift slices and use that knowledge to better detect filaments at less populated redshift slices. This proves especially useful when extending the reconstruction to higher redshift slices. We show that, according to the metrics above, this approach also provides more uniform results across redshifts.
                \item We study the correlation between our cosmic filament catalogue and external galaxy cluster catalogues from Planck and ACT as part of the validation of our catalogue. These galaxy cluster catalogues are built using the Sunyaev-Zel'dovich effect, a completely independent physical probe. We show a very strong correlation with both catalogues, up to more than $20\sigma$ compared to random points.
                \item We study the filament persistence with a realistic example by comparing the filaments found with older data (BOSS) and with all data (BOSS+eBOSS) at redshift slices between $z=0.6$ and $z=0.7$. We show that excluding the new data (up to $60\%$ of total points) only results in a loss of less than $25\%$ of the reconstructed cosmic web.
                \item Although the algorithm is not designed to isolate single filaments, we performed a first analysis of the filament length distribution assuming that filament endpoints are in the intersections. We find similar trends to those obtained in other works \citep{malavasi2020, galarragaespinosa2020}
        \end{itemize}
        
        The catalogue is publicly available at \href{https://www.javiercarron.com/catalogue}{javiercarron.com/catalogue}, along with Python codes that can be used to perform standard operations.
        
        We hope that this new catalogue opens the door to more complete studies of the cosmic web. In particular, we are currently working on the extension and validation of previously found correlations between filaments and weak lensing of the cosmic microwave background and the Sunyaev-Zeldovich effect in order to constrain mass profiles and their hot gas content \citep{he2018, tanimura2020}. The methodology in this work can also be applied to N-body simulations in order to assess the feasibility of using cosmic filaments as cosmological probes. Finally, we believe that the tomographic reconstruction of the cosmic web will be extremely valuable for upcoming wide photometric galaxy surveys like those expected from Euclid and LSST, for which three-dimensional reconstruction is more difficult.

        \section*{Acknowledgements}
        
        We thank Yen-Chi Chen for the valuable discussions regarding the SCMS algorithm and the comparison of the catalogues. We are grateful to the referee for very useful comments and suggestions, which helped improve the manuscript.
        
        We have developed the code used in this work using \textit{Python 3.8} (available at \href{https://www.python.org/}{python.org}) with the JupyterLab environment \citep{jupyter}, and the following packages. For numerical computations, we use \textit{NumPy} \citep{numpy}, \textit{SciPy} \citep{scipy}, and \textit{Astropy} \citep{astropy}. For treatment of the spherical maps, we use \textit{healpy} \citep{gorski2005,healpy} and \textit{MTNeedlet} \citep{carronduque2019}. For treatment of tables and databases, we use \textit{pandas} \citep{pandas}. For parallelization of the code, we use \textit{Ray} \citep{ray}. For visualisation, we use \textit{Matplotlib} \citep{matplotlib} and \textit{seaborn} \citep{seaborn}. We thank the developers of these pieces of software and their contributors.
        
        We acknowledge support from INFN through the InDark initiative. This work was also supported by ASI/COSMOS grant n. 2016-24-H.0 and ASI/LiteBIRD grant n. 2020-9-HH.0. M.M. is supported by the program for young researchers ``Rita Levi Montalcini" year 2015. D.M. acknowledges support from the MIUR Excellence Project awarded to the Department of Mathematics, Universit\`{a} di Roma Tor Vergata, CUP E83C18000100006.
        
        Funding for the Sloan Digital Sky Survey IV has been provided by the Alfred P. Sloan Foundation, the U.S. Department of Energy Office of Science, and the Participating Institutions. SDSS acknowledges support and resources from the Center for High-Performance Computing at the University of Utah. The SDSS web site is \url{www.sdss.org}.
        
        SDSS is managed by the Astrophysical Research Consortium for the Participating Institutions of the SDSS Collaboration including the Brazilian Participation Group, the Carnegie Institution for Science, Carnegie Mellon University, Center for Astrophysics | Harvard \& Smithsonian (CfA), the Chilean Participation Group, the French Participation Group, Instituto de Astrofísica de Canarias, The Johns Hopkins University, Kavli Institute for the Physics and Mathematics of the Universe (IPMU) / University of Tokyo, the Korean Participation Group, Lawrence Berkeley National Laboratory, Leibniz Institut für Astrophysik Potsdam (AIP), Max-Planck-Institut für Astronomie (MPIA Heidelberg), Max-Planck-Institut für Astrophysik (MPA Garching), Max-Planck-Institut für Extraterrestrische Physik (MPE), National Astronomical Observatories of China, New Mexico State University, New York University, University of Notre Dame, Observatório Nacional / MCTI, The Ohio State University, Pennsylvania State University, Shanghai Astronomical Observatory, United Kingdom Participation Group, Universidad Nacional Autónoma de México, University of Arizona, University of Colorado Boulder, University of Oxford, University of Portsmouth, University of Utah, University of Virginia, University of Washington, University of Wisconsin, Vanderbilt University, and Yale University.
        
        Based on observations obtained with Planck (\url{http://www.esa.int/Planck}), an ESA science mission with instruments and contributions directly funded by ESA Member States, NASA, and Canada.
        
        This work used the public data produced by the Atacama Cosmology Telescope (\url{https://act.princeton.edu}).

        \bibliographystyle{aa} 
        \bibliography{bibtex} 

        \appendix
        \section{Training of the gradient boosting method}
        \label{ap:train}
        As explained in \Cref{ss:combine_B}, we use a gradient boosting algorithm as a part of Method B. The input corresponds to five quantities computed at four scales each, for each pixel. This means that the input corresponds to $20$ features and as many points as pixels in the observed region. The algorithm uses the value of the quantities at each pixel to predict the typical distance of a filament from a pixel with certain characteristics. 
        
        In order to create the training maps, we proceed as follows. We select an intermediate value of redshift and obtain the estimated uncertainty map for each of the four smoothing scales mentioned above. The uncertainty map is obtained by applying \Cref{alg:2} to all pixels in the observed sky. This map represents the typical distance from each pixel to filaments of the bootstrapping simulation, when using the algorithm at a certain smoothing scale. Let $x_{n,s}$ be the value of this typical distance at the pixel $n=1...N$ and smoothing scale labelled by $s=1...S$, where $S=4$ in our case. We combine all the maps into $\bar{x}_n$ with the following expression:
        \begin{equation}
                \bar{x}_n = \left(\frac{1}{S} \sum_{s=1}^{S}  \frac{x_{n,s}}{\frac{1}{N}\sum_{n=1}^{N} x_{n,s}}\right) \cdot \left(\frac{1}{S N}\sum_{n, s} x_{n,s}\right)
        .\end{equation}
        
        This expression can be understood as an arithmetic average of the four scales, where each size is normalised to unit mean in order to give the same importance to all scales. Then, the result of the average is denormalised to the original scale by multiplying it with the global average, so that the final result $\bar{x}_n$ is given in units of degrees, as in the original $x_{n,s}$. More complex expressions are possible, such as giving different weights to different scales; however, in practice they do not significantly affect the filament reconstruction.
        
        In principle, this procedure could be done for all $284$ redshift slices instead of using it to train a gradient boosting algorithm. However, this would be computationally unfeasible, because it requires the application of \Cref{alg:1} a total of $401$ times per redshift slice ($100$ different bootstrap simulation times $4$ scales, and $1$ final computation in order to obtain the filaments), plus the computation of millions of pair-wise distances between points on the sphere. By training a machine learning algorithm, we reduce this computation to only $5$ per slice ($1$ times $4$ scales in order to compute the input values and $1$ final computation in order to obtain the filaments). Training the algorithm has other advantages, such as a more robust detection at higher redshift, with less data.

        In order to have training data representing all the redshift region, we repeat the previous computation for a redshift slice in each of the different sets of data. In practice, we select the slices starting at $z=0.239$ for LOWZ, $z=0.570$ for CMASS, $0.803$ for LRG, and $z=1.394$ for QSO. Then, we apply the trained gradient boosting to all slices. We note that the results for a single slice interpolate very well within the same survey and even to the different surveys at different redshift values. However, we use a slice from each survey as this approach is more general and could include particular effects from all surveys, making it more robust.
        
        The \textit{mean squared error} of the predictor is $0.047$ $\deg$ on the test dataset. We note that the exact predicted values are not critical, as we are interested in the ridges of the map. The results concerning the filaments reconstructed with this method are reported in \Cref{s:results} under the label of Method B.

        \section{LRG and QSO in the training sample}
        \label{ap:lrg-qso}
        
        We used a training sample representing all the different surveys we are using. However, we know that LRGs and QSOs are different tracers of the large-scale structure. In particular, QSOs are known to present a higher value of the bias \citep{laurent2017}, meaning that they tend to be more clustered along dark matter overdensities than LRGs.
        
        In principle, this means that the training sample could be inhomogeneous and the training could be affected, or even impossible. On the other hand, the fact that QSOs are better tracers of the cosmic web could be balanced by the lower number of them. Additionally, the underlying cosmic web traced by both populations is the same, so the algorithm may be flexible enough to learn some characteristics linked to the web itself rather than its tracers.
        
        In order to test possible effects caused by the differences between LRGs and QSOs, we train the Gradient Boosting step (as explained in \Cref{ss:combine_B}) on both populations independently. For LRGs, we restrict the training sample to the LOWZ and CMASS training slices ($z=0.239$ and $z=0.570$), while for QSOs, we only use the eBOSS QSO training sample ($z=1.394$).
        
\begin{figure*}
                \centering
                \includegraphics[width=0.7\textwidth]{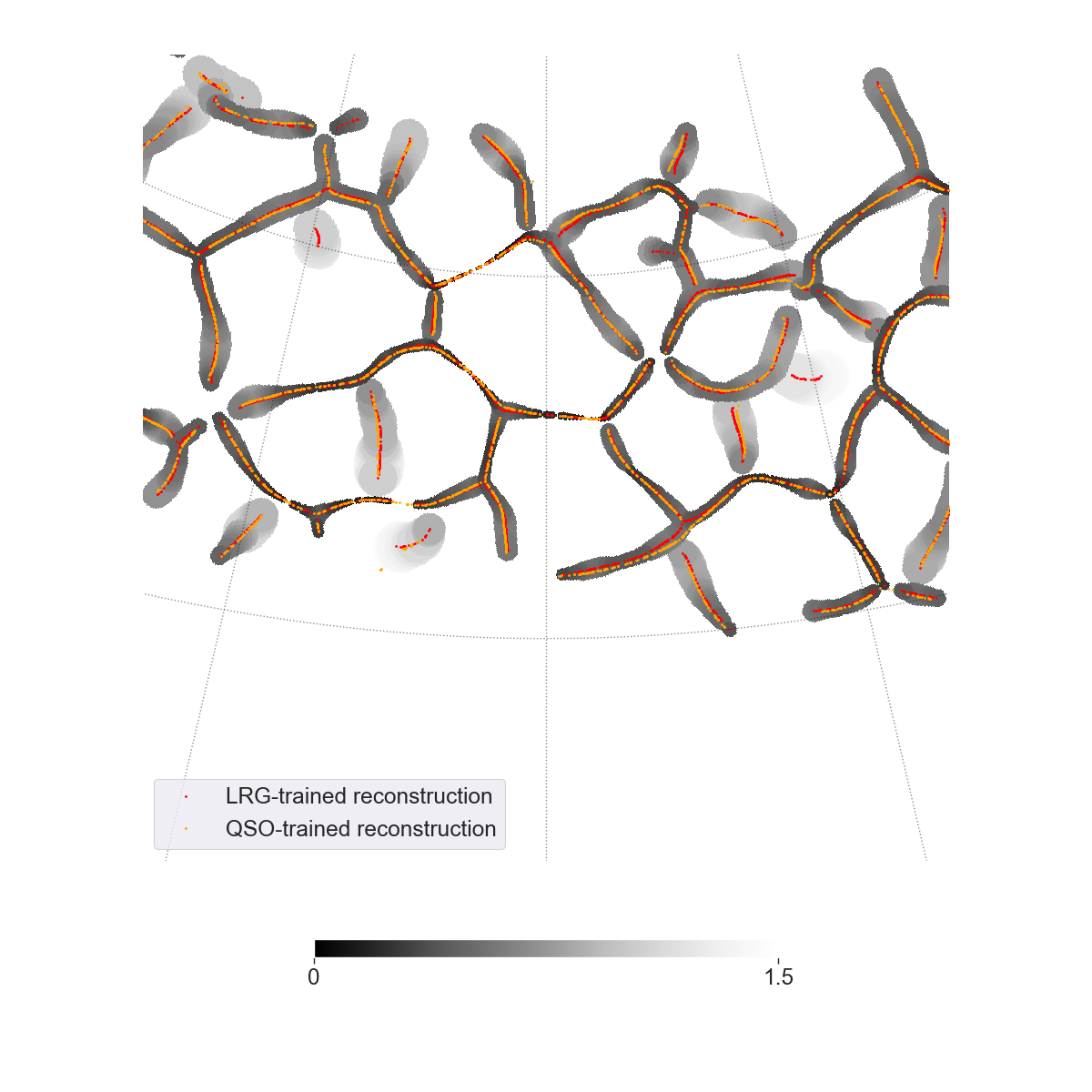} 
                \caption{Comparison of the reconstructions obtained by training the algorithm only with LRGs (red) and only with QSOs (orange). The coloured region is the confidence region of the LRG-reconstructed filaments. The figure is a gnomic projection centred at $RA=\ang{180}$, $dec=\ang{40}$ with a side size of \ang{45}. The data used in this example correspond to data at $z=0.803$.
                        \label{f:trainvalidation}}
        \end{figure*}

        After training the algorithm on both populations independently, we apply it to a sample slice at $z=0.803$. We can see the filaments obtained in both cases in \Cref{f:trainvalidation}: the filament network reconstruction is almost identical in both cases, and they are well inside the confidence region of each other. We checked that this result is consistent in the whole redshift range.

        We also analyse the similarity between the Gradient Boosting algorithm trained in these two sets. A way to do this is to give them the same input and analysing the similarity among the results. We feed them with the real inputs for several redshift slices and we see a very high correlation between both outputs ($R=0.97$). It has to be noted that the training on QSOs yields slightly higher predictions than when training on LRGs, perhaps because of the larger uncertainties driven by the lower number of QSOs. This means that the uncertainty estimates of the detections are slightly larger if we only use QSOs in the training ($\sim$\ang{;10;} larger); however, the positions of the detected filaments, presents no difference.

        \section{Filament catalogue cleaning}
        \label{ap:clean}
        
        Once the filaments have been obtained by either of the two methods explained in \Cref{ss:combine}, we apply a series of filters to remove spurious or unclear detections. All the thresholds adopted in this work are very conservative in order to remove only the outliers and problematic regions, preserving a very large sample. However, we recommend imposing more aggressive thresholds depending on the specific application of the catalogue. An especially useful threshold may be in the estimated uncertainty of the detections. These filters can easily be performed with the code provided with the catalogue at \href{https://www.javiercarron.com/catalogue}{javiercarron.com/catalogue}.
        
        \begin{itemize}
                \item Initial density estimate: step 6 of \Cref{alg:1} introduces an initial set of points ${y_j}$ that are used to start the iterative algorithm. In practise, we remove the points with a density estimate lower than a threshold $K$. This is a common criterion in filament-searching algorithms \citep{cautun2013,chen2015} as it helps with the stability of the results and minimises spurious detections. In our case, we define $K$ to be the $10\%$ percentile of the value of the density estimate in the region of interest.
                \item Border effects: some artefacts may arise in the areas of the sky near the border of the surveyed regions. In order to remove these areas, we establish a buffer of $~2\deg$: the reported filaments are the detections that are further than this distance from a border.
                \item Threshold in estimated uncertainty: the location of spurious filaments may vary significantly if the initial set of points vary slightly. This is estimated through \Cref{alg:2}. We remove detections whose estimated uncertainty is in the top $2\%$ of the slice (corresponding to an estimated uncertainty of \SIrange{1.0}{1.5}{\deg}). We note that most applications of the catalogue will require a more strict threshold or a threshold in units of proper distance.
                \item Isolated points: the algorithms sometimes detect spurious ridges that consist in a few points, not connected to any other structure. By contrast, typical filaments are detected by hundreds or thousands of points. This effect is rare but can be easily removed: detections are only reported if they have other detections nearby. In practice, we consider a coarser pixelisation (with $N_{side}=128$) and count the detections in each pixel; in order to be reported, a point must be in a pixel with at least $15$ detections or adjacent to one of such pixels.
        \end{itemize}
        
        These filters are applied consecutively in the listed order.

        The average number of points excluded by each filter in each block can be seen in \Cref{t:filters}. Within Blocks, there are only weak trends with redshift, as every filter has a similar effect regardless of redshift.
        
        \begin{table*}
                \caption{Average number of points excluded due to each of the preprocessing filters applied in order, see text.
                        \label{t:filters}}
                \centering
                \begin{tabular}{ccccccc}
                        Block & Total points & Initial density & Border & Uncertainty & Isolation & Remaining \vspace{1pt}\\
                        \hline
                        Catalogue A &&&&&& \\
                        Block 1 & $194155$ & $33671 \pm 3104$ & $20404 \pm 2914$ & $2802\pm39$ & $1046\pm350$ & $136231\pm2093$ \\
                        &  & ($17.34\%$) & ($10.51\%$) & ($1.44\%$) & ($0.54\%$) & ($70.17\%$) \vspace{1pt}\\
                        Block 2 & $209403$ & $38764\pm1334$ & $10440\pm1738$ & $3204\pm18$ & $2191\pm508$ & $154805\pm1093$ \\
                        &  & ($18.51\%$) & ($4.99\%$) & ($1.53\%$) & ($1.05\%$) & ($73.93\%$) \vspace{1pt}\\
                        Block 3 & $68628$ & $17891\pm1397$ & $4262\pm1779$ & $930\pm15$ & $407\pm159$ & $45138\pm817$ \\
                        &  & ($26.07\%$) & ($6.21\%$) & ($1.36\%$) & ($0.59\%$) & ($65.77\%$) \vspace{1pt}\\
                        \hline
                        Catalogue B &&&&&& \\
                        Block 1 & $194155$ & $62552 \pm 2$ & $4693 \pm 757$ & $2539\pm15$ & $1297\pm192$ & $123075\pm793$ \\
                        &  & ($32.22\%$) & ($2.42\%$) & ($1.31\%$) & ($0.67\%$) & ($63.39\%$) \vspace{1pt}\\
                        Block 2 & $209403$ & $54556\pm2$ & $2528\pm362$ & $3047\pm7$ & $1459\pm208$ & $147814\pm363$ \\
                        &  & ($26.05\%$) & ($1.21\%$) & ($1.45\%$) & ($0.70\%$) & ($70.59\%$) \vspace{1pt}\\
                        Block 3 & $68628$ & $24851\pm1$ & $1339\pm280$ & $849\pm6$ & $701\pm159$ & $40887\pm328$ \\
                        &  & ($36.21\%$) & ($1.95\%$) & ($1.24\%$) & ($1.02\%$) & ($59.58\%$) \vspace{1pt}\\
                \end{tabular}
            \tablefoot{The uncertainty is the standard deviation of the slices in a block. The percentages refer to the proportion of total points.}
        \end{table*}

        Additionally, we reduce the size of the catalogue by removing most points. The original search of filaments is done with a uniform grid in the sky (the grid $\{y_j\}$ in \Cref{alg:1}); in particular, we use the centres of the pixels in the HEALPix scheme with $nside=256$. For example, this translates into more than \num{210000} points per redshift slice for Block 2. Considering all the $284$ redshift slices, we have more $34.9$ million points (more than $24.2$ after the filters above). However, this set of points is highly redundant, because each filament is detected by hundreds or thousands of points. It is possible to reduce the size of the catalogue without losing any information. We sample $20\%$ of this initial set of points to obtain \num{4862353} points for Catalogue A and \num{4460054} points for Catalogue B. This is performed as a final step after all other filters have been applied.

\end{document}